\documentclass[journal=jpcafh,manuscript=article]{achemso}
\SectionNumbersOn  

\usepackage{amsmath}
\usepackage{amssymb}
\usepackage{physics}
\usepackage{graphicx} 
\usepackage{enumerate}
\usepackage[inline]{enumitem}
\usepackage{bm}
\usepackage{tcolorbox}
\usepackage{xcolor}
\usepackage[normalem]{ulem}
\usepackage{tikz}

\usepackage{hyperref}
\hypersetup{hidelinks,
  colorlinks=true,
  allcolors=blue,
  pdfstartview=Fit,
  breaklinks=true
}

\newcommand{\im}{\textup{i}}

\author{Rui-Hao Bi}
\affiliation{Department of Chemistry, School of Science and Research Center for Industries of the Future, Westlake University, Hangzhou, Zhejiang 310024, China}
\alsoaffiliation{Institute of Natural Sciences, Westlake Institute for Advanced Study, Hangzhou, Zhejiang 310024, China}

\author{Chongxiao Zhao}
\affiliation{Department of Chemistry, School of Science and Research Center for Industries of the Future, Westlake University, Hangzhou, Zhejiang 310024, China}
\alsoaffiliation{Institute of Natural Sciences, Westlake Institute for Advanced Study, Hangzhou, Zhejiang 310024, China}
\alsoaffiliation{Department of Chemistry, Zhejiang University, Hangzhou 310027, China}

\author{Ruixin Sun}
\affiliation{Department of Physics, School of Science and Research Center for Industries of the Future, Westlake University, Hangzhou, Zhejiang, 310024, China}
\alsoaffiliation{Institute of Natural Sciences, Westlake Institute for Advanced Study, Hangzhou, Zhejiang 310024, China}

\author{Wenjie Dou}%
\email{douwenjie@westlake.edu.cn} 
\affiliation{Department of Chemistry, School of Science and Research Center for Industries of the Future, Westlake University, Hangzhou, Zhejiang 310024, China}
\alsoaffiliation{Institute of Natural Sciences, Westlake Institute for Advanced Study, Hangzhou, Zhejiang 310024, China}
\alsoaffiliation{Department of Physics, School of Science and Research Center for Industries of the Future, Westlake University, Hangzhou, Zhejiang, 310024, China}
\alsoaffiliation{Key Laboratory for Quantum Materials of Zhejiang Province, Department of Physics, School of Science and Research Center for Industries of the Future, Westlake University, Hangzhou, Zhejiang 310024, China}

\title{Assessing the Performance of CC2 for Excited-State Dynamics: A Benchmark Study with Pyrazine}

\begin{document}

\begin{abstract}
During the ultrafast electronic relaxation of pyrazine, the bright state undergoes rapid decay accompanied by coupled electronic and nuclear motion. We investigate the vibrational modes associated with this process using trajectory surface hopping simulations. To enable accurate \emph{ab initio} excited-state dynamics, we apply CC2 analytical gradients and nonadiabatic coupling vectors within the Q-Chem package. Using both a reduced vibronic coupling model and full-dimensional simulations, we demonstrate that CC2 predicts that the dark $A_\text{1u}$ state actively participates in the initial relaxation and the subsequent dynamics. The full-dimensional simulations yield a $B_\text{2u}$ decay time of 25--26 fs, in good agreement with the experimental measurement of $22\pm3$ fs. Furthermore, we identify the $Q_\text{8a}$ and $Q_\text{9a}$ modes as dominant tuning coordinates whose motions are phase-locked to the population oscillations between the $A_{1u}$ and $B_{3u}$ states.
\end{abstract}

\begin{tocentry}
  \includegraphics[width=3.25in,height=1.75in,keepaspectratio]{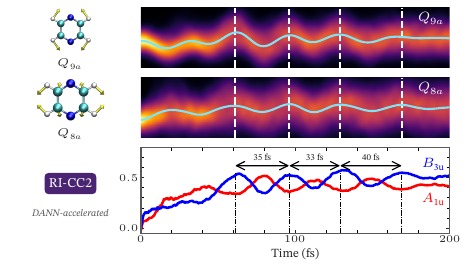}
\end{tocentry}

\maketitle

\section{Introduction}
After photoexcitation, the electronic populations of a molecule rapidly evolve among a manifold of electronic states, strongly coupled to the molecular motions. This coupled dynamics gives rise to rich photophysical and photochemical phenomena such as vision and photocatalysis. The pursuit of a detailed understanding of these processes has spawned numerous active fields in both experiment and theory. To gain straightforward, microscopic insights, \emph{ab initio} simulations have become indispensable. Among these, trajectory surface hopping (TSH) has emerged as one of the most widely used approaches for modeling nonadiabatic dynamics, due to its favorable balance between efficiency and accuracy~\cite{tully_jcp1990_fssh, subotnik_annurev2016_surfacehopping, crespo-otero_chemrev2018_mqc}.

To perform TSH simulations, the second-order approximate coupled-cluster singles and doubles (CC2\cite{christiansen_jcp1995_cc2_gs,christiansen_jcp1996_cc2_excitation}) strikes a favorable balance between accuracy and computational cost. However, its adoption in nonadiabatic dynamics has been hindered by the scarcity of analytical gradients and nonadiabatic coupling vectors (NACVs) in standard software packages~\cite{furche_Turbomole2014}, despite significant methodological
progress~\cite{tajtiAnalyticEvaluationNonadiabatic2009,farajiCalculationsNonadiabaticCouplings2018,kjonstadBiorthonormalFormalismNonadiabatic2021,chatterjeeAnalyticEvaluationNonadiabatic2023,kjonstadCommunicationNonadiabaticDerivative2023,kjonstad_jctc2024_nac,rossiGeneralizedCoupledCluster2025,stollSimilarityConstrainedCC22025}. Furthermore, numerical instabilities near conical intersections---such as those reported in 9H-adenine~\cite{plasser_jctc2014_cc2_crash}---have limited practical
applications. Recently, Angelico and coworkers have shown that similarity-constrained coupled cluster (EOM-SCCSD\cite{stollSimilarityConstrainedCC22025,angelico2026scc}) methods can avoid such instabilities when quasidegenerate states are encountered in nonadiabatic dynamics. In this work, we demonstrate the application of our recently implemented analytical gradients and nonadiabatic coupling vectors (NACVs) for RI-CC2---available within the Q-Chem
package~\cite{zhao2024_sri_es, zhao2024_sri_gs, zhao2025_srifosc, zhao2025_sri_grad_nac}---to the benchmark pyrazine system. Although this standard RI-CC2 implementation remains susceptible to known numerical instabilities near conical intersections, such as complex excitation energies and incorrect conical intersection dimensionalities~\cite{Thomas17112021}, our simulations on pyrazine provide a practical test case for evaluating the performance and limitations of this approach in nonadiabatic dynamics.

Pyrazine is a classic model system for studying radiationless transitions in the intermediate regime~\cite{byrne1965InternalConversionAromatic,robinson_jcp1967_isc_gas,kommandeur_annrev1987_pyrazine}. Crucially, the dynamics in this system are dominated by a few key vibrational modes strongly coupled to the electronic manifold, making it an ideal testbed for isolating the vibrational motions that drive internal conversion. Although CC2 is a single-reference method, it is suitable for this study because the excited states do not interact with the ground state during the ultrafast internal conversion process~\cite{kohn_jcp2007_cc4ci, xie2019_assessing, liMachineLearningAccelerated2023a}.

The theoretical effort began with two-state vibronic coupling models that were progressively refined to describe the coupled $B_\text{2u}(\pi\pi^{*})$ and $B_\text{3u}(n\pi^{*})$ states~\cite{schneider1988S1S2CI, domcke1993CI_S0S2_CASSCF, sobolewski1993CI_S0S2_MRCI, shiozaki_pccp2013_ms_lvc}. These models included several key tuning modes ($Q_\text{1}$, $Q_\text{6a}$) and the $Q_\text{10a}$ coupling mode, enabling early simulations of absorption spectra~\cite{schneider1988S1S2CI, stock_jcp1989_pp, stock1993_pop_pumpprobe, worth1996mctdh_bath, raab1999spectra_full24modes}, pump--probe spectroscopies~\cite{stock_jcp1989_pp, stock1993_pop_pumpprobe}, resonance Raman and fluorescence~\cite{stockTheoryResonanceRaman1990}, ionization spectroscopy~\cite{seel1991tdpes_theory, hahn2001_tdpes_theory, suzuki2003ang_tdpes_theory}, and optimal control~\cite{christopher2005coherent_control, christopher2006active_control_full, grinev_jpb2015_coherent_control}. Later quantum dynamics studies treated the remaining modes either as a bath~\cite{worth1996mctdh_bath} or explicitly in full dimensionality~\cite{raab1999spectra_full24modes, puzari_jcp2006_24modes}, and solvent effects on internal conversion were also investigated~\cite{burghardt2006effect_of_bath}. Experimental advances soon followed, with time-resolved photoelectron spectroscopy (TRPES) reporting an internal conversion timescale of $20\pm10$ fs~\cite{stert2000tdpes_spec}, in good agreement with earlier simulations~\cite{seel1991tdpes_theory}. Improved temporal resolution was subsequently achieved with time-resolved photoelectron imaging (TRPEI), yielding a timescale of $22\pm3$ fs~\cite{suzuki2006fs_trpei, horio2009trpei, suzuki_jcp2010_trpei}. These early studies of pyrazine collectively advanced the development of both theoretical methods and experimental techniques, establishing the $B_\text{2u} \rightarrow B_\text{3u}$ pathway and the importance of modes such as $Q_\text{6a}$ and $Q_\text{10a}$.

While these foundational studies captured the primary decay channel, more recent investigations have uncovered a richer and more debated landscape centered on the involvement of the dark $A_\text{1u}$ state and the precise assignment of the vibrational modes responsible for the observed quantum beats. The potential involvement of the $A_\text{1u}$ state was first noted by Werner \emph{et al.} in \emph{ab initio} on-the-fly TSH simulations using time-dependent density functional theory (TD-DFT)~\cite{werner_cp2008_tddft_b3lyp}. They subsequently employed TD-DFT to simulate TRPES~\cite{werner_jcp2010_dft_tdpes} and TRPEI~\cite{tomaselloExploringUltrafastDynamics2014} spectra, identifying the $Q_\text{6a}$ mode as the primary driver of the quantum beats in TRPES. These findings inspired a reparameterization of a three-state vibronic coupling model that explicitly includes the $A_\text{1u}$ state, performed at the extended multireference perturbation theory (XMCQDPT2) level~\cite{sala_pccp2014_RoleLowlyingDark, sala_pccp_qd2015}. This model indicated that the $A_\text{1u}$ state actively participates in the internal conversion and predicted that the $Q_\text{8a}$ mode is associated with the population oscillation between $A_\text{1u}$ and $B_\text{3u}$. These interpretations were later supported by \emph{ab initio} on-the-fly dynamics at the algebraic diagrammatic construction ADC(2) level~\cite{xie2019_assessing}. Time-resolved NEXAFS simulations by Tsuru \emph{et al.}~\cite{tsuru_jcp2019_nexafs}, based on the same vibronic-coupling model, further identified a minor $\approx$30 fs population fluctuation assigned to Q$_\text{9a}$, in addition to the dominant $\approx$60 fs Q$_\text{6a}$ oscillation. Additional theoretical progress includes studies of pyrazine internal conversion in the condensed phase~\cite{vogt_jctc2025_condphase}, quasiclassical doorway-window simulations of transient-absorption pump-probe signals~\cite{gelinInitioSurfaceHoppingSimulation2021, guanQuasiclassicalDoorwayWindow2025, liFinetuningLaserpulseFrequencies2025}, and machine learning--accelerated nonadiabatic dynamics extending to the picosecond timescale~\cite{buzsaki_chemrxiv2026}.

The role of the $A_\text{1u}$ state has sparked considerable debate between experimentalists and theoreticians. Horio \emph{et al.} did not observe signatures of the $A_\text{1u}$ state in their vacuum ultraviolet TRPEI experiments~\cite{horio_jcp2016_noA1u}. Subsequently, Mignolet \emph{et al.} simulated the same TRPEI experiments at the multireference configuration interaction with singles and doubles (MRCISD) level and found the predicted $A_\text{1u}$ signal to be markedly different from the
experimental results~\cite{mignolet_cp2018_noa1u}. Kanno \emph{et al.} also simulated the wavepacket dynamics of pyrazine without reaching a definitive conclusion regarding the $A_\text{1u}$ state~\cite{kanno_jcp2021_noa1u}. However, Pit\v{s}a \emph{et al.} argued that the TRPEI signals of the $A_\text{1u}$ and $B_\text{3u}$ states are effectively indistinguishable from one another~\cite{pitesa_jctc2021_indistingushable}. This debate was later addressed by an X-ray transient absorption study by
Scutelnic \emph{et al.}, which directly observed the $A_\text{1u}$ state~\cite{scutelnic_natcomm2021_xray}. Recently, Karashima \emph{et al.} conducted a TRPES study with a time resolution of 13.3 fs, revealing that the internal conversion between the $A_\text{1u}$ and $B_\text{3u}$ states proceeds through oscillatory population exchange driven by the $Q_\text{6a}$ mode, as identified through Fourier analysis of the TRPES signal~\cite{karashima_jacs2024_vibmotion}. Chang \emph{et al.} combined X-ray absorption spectroscopy at the nitrogen K-edge with surface hopping simulations, revealing that the population transfer accompanies cyclic rearrangement of the electronic structure as the nuclear wavepacket encircles the crossing along the $Q_\text{8a}$ tuning mode and the $Q_\text{8b}$ coupling mode.\cite{Chang2025}

The role of the A$_\text{1u}$ state in theoretical studies is complicated by the fact that its state ordering is highly method-dependent. Luke \emph{et al.} comprehensively assessed the state ordering at the Franck--Condon geometry using different electronic structure methods and demonstrated that whether the A$_\text{1u}$ state lies below or above the B$_\text{2u}$ state depends on the treatment of dynamic correlation \cite{state_ordering2026}. In this work, we assess how the CC2 method predicts this state ordering.

Despite these advances, a consistent microscopic picture connecting the electronic state ordering with specific vibrational signatures---particularly reconciling the predictions for $Q_\text{6a}$ versus $Q_\text{8a}$ activity---has remained elusive. In this work, we leverage our RI-CC2 surface-hopping framework to characterize the coupled electron--nuclear motions underlying the internal conversion of pyrazine. Building on the rich experimental and theoretical landscape surrounding pyrazine, we specifically investigate the role of the dark $A_\text{1u}$ state and the vibrational modes associated with the correlated population transfer. Section~\ref{sec:methods} outlines the theoretical methods, including RI-CC2 (Sec.~\ref{subsec:ricc2}), the vibronic coupling model (Sec.~\ref{subsec:vc_model}), trajectory surface hopping and diabatization (Sec.~\ref{subsec:fssh}), and the DANN machine-learning force field (Sec.~\ref{subsec:dann}). Computational details are provided in Sec.~\ref{sec:comp_details}. Our simulation results are presented in Sec.~\ref{sec:sim_res}, followed by concluding remarks in Sec.~\ref{sec:conclusion}.

\section{\label{sec:methods}Methods}
\subsection{\label{subsec:ricc2}Excited state calculations with RI-CC2}
At an arbitrary molecular geometry $\bm{R}$, the ground and excited singlet states of pyrazine are described by the Schrödinger equation $\hat{H}(\bm{R}) \ket{\text{S}_n(\bm{R})} = E_n(\bm{R}) \ket{\text{S}_n(\bm{R})}, $ where $\ket{\text{S}_n(\bm{R})}$ and $E_n(\bm{R})$ denote the $n$-th singlet state and its corresponding energy. To study the nonadiabatic dynamics of pyrazine, the RI-CC2 method was employed~\cite{christiansen_jcp1995_cc2_gs,christiansen_jcp1996_cc2_excitation,vahtras_cpl1993_ri,feyereise_cpl1993_ri}. The ground state $\text{S}_0$ is described by the coupled-cluster ansatz using a single Hartree--Fock reference determinant\cite{christiansen_jcp1995_cc2_gs,zhao2024_sri_gs}. Excitation energies $E_{n\geq1}$ are obtained from coupled-cluster linear response theory\cite{christiansen_jcp1996_cc2_excitation,zhao2024_sri_es}. Analytical energy gradients for both the ground state ($\partial E_0/\partial\bm{R}$) and excited states ($\partial E_{n\geq1}/\partial\bm{R}$) are computed using the Lagrangian formalism\cite{AnalyticalCalculationGeometrical1988,christiansen_1998_gradient,hattig_jcp2003_gsgradient,kohn_jcp2003_exgradient,ledermuller_jcp2014_laplace,zhao2025_srifosc,zhao2025_sri_grad_nac}. 

Transition properties in this work are calculated using RI-CC2 within the linear response formalism\cite{christiansen_jcp1998_transition,hattig_jcp2002_transition_ricc2,kohn_jcp2007_cc4ci,kjonstad_jctc2024_nac,zhao2025_srifosc,zhao2024_sri_es}. Due to the non-Hermitian nature of coupled-cluster theory, the left- and right-transition dipole moments are not identical:
\begin{equation}\label{eqn:dipoles}
    \bm{\mu}_{n\gets0}(\bm{R}) = \mel{\text{S}_n(\bm{R})}{\hat{\mu}}{\text{S}_0(\bm{R})}, \quad \bm{\mu}_{0 \gets n} = \mel{\text{S}_0(\bm{R})}{\hat{\mu}}{\text{S}_n(\bm{R})}.
\end{equation}
The oscillator strength, however, can be obtained from these dipoles as
\begin{equation}
    f_{n\gets0} = \frac{2}{3}(E_n-E_0) \bm{\mu}_{0\gets{n}} \cdot \bm{\mu}_{{n}\gets 0}.
\end{equation}
Similarly, the NACVs in RI-CC2,
\begin{equation}
    \bm{d}_{mn}(\bm{R}) = \braket{\text{S}_m(\bm{R})}{\grad\text{S}_n(\bm{R})},
\end{equation}
are also not antisymmetric, such that $\bm{d}_{mn} \neq -\bm{d}_{nm}$. In this work, we only calculate the NACVs $\bm{d}_{mn}$ with $m>n$.

We note that the standard RI-CC2 implementation employed here is known to exhibit numerical instabilities in certain regions of nuclear configuration space, such as complex-valued excitation energies and incorrect conical intersection dimensionalities\cite{crespo-otero_chemrev2018_mqc,plasser_jctc2014_cc2_crash,Thomas17112021,stollSimilarityConstrainedCC22025}. These limitations are inherent to the linear-response CC2 framework and are not resolved in the present implementation. Near the defective region, the excitation-energy surfaces become square-root-like, and the gradients and nonadiabatic couplings become discontinuous or divergent; this artifact can crash the simulation or alter the nuclear dynamics.

\subsection{\label{subsec:vc_model}Vibronic coupling model}
The ultrafast internal conversion process of pyrazine can be described by a low-dimensional vibronic coupling (VC) model~\cite{schneider1988S1S2CI,worth1996mctdh_bath,raab1999spectra_full24modes,sala_pccp2014_RoleLowlyingDark}. Beyond providing an intuitive visualization of the potential-energy landscape (Figure~\ref{fig:pes}), this reduced model offers a controlled benchmark for the trajectory surface hopping protocol, as it allows for a direct comparison with exact quantum dynamics in the same reduced space.

Following Sala \emph{et al.}~\cite{sala_pccp2014_RoleLowlyingDark} and Xie \emph{et al.}~\cite{xie2019_assessing}, we consider the three lowest excited states of pyrazine---the weakly bright $B_\text{3u}$ state, the dark $A_\text{1u}$ state, and the bright $B_\text{2u}$ state---and construct the following VC model as a function of the dimensionless normal mode coordinates $\bm{Q}$:
\begin{equation}
    H_\text{vc}(\bm{Q}) = H_0(\bm{Q}) + V(\bm{Q}).
\end{equation}
Here, the reference Hamiltonian $H_0$ represents the ground state within the harmonic approximation:
\begin{equation}\label{eqn:vc_ref}
    H_0(\bm{Q}) = \sum_{i=1}\frac{\hbar\omega_i}{2} (P_i^2 + Q_i^2) I, 
\end{equation} 
where $P_i$ and $Q_i$ are the momentum and position operators for the $i$-th vibrational mode with frequency $\omega_i$, and $I$ is the identity operator in the space of the three diabatic states. The diagonal elements of the potential matrix $V$ include linear and quadratic couplings:
\begin{equation}\label{eqn:vc_diag}
    V_{nn}(\bm{Q}) = E_n^\text{eq} + \sum_i  \left(\kappa_i^{n} Q_i + \gamma_i^{n} Q_i^2\right),
\end{equation}
where $E_n^\text{eq}$ is the excitation energy of the $n$-th singlet state at the reference (ground-state equilibrium) geometry, and $\kappa_i^{(n)}$ and $\gamma_i^{(n)}$ are the linear and quadratic coupling coefficients, respectively. For the off-diagonal couplings, we consider only linear terms:
\begin{equation}\label{eqn:vc_offdiag}
    V_{nn'} = \sum_{j} \lambda_j^{nn'} Q_j,
\end{equation}
where $\lambda_j^{(nn')}$ is the linear coupling coefficient between diabatic states $n$ and $n'$ for the $j$-th mode.

The linear coupling coefficients $\kappa_i^{(n)}$ and $\lambda_i^{(nn')}$ are parameterized from RI-CC2 calculations following the standard procedure described in Ref.~\cite{plasser_pccp2019_parametrization}:
\begin{gather}
    \kappa_i^{n} = \sqrt{\frac{\hbar}{\omega_i}} \sum_{\alpha} \frac{K_{\alpha i}}{\sqrt{M_{\alpha}}} \eval{\pdv{E_n}{R^{\alpha}}}_{\bm{R}=\bm{R}^\text{eq}}, \label{eqn:kappa}\\
    \lambda_i^{nn'} = \sqrt{\frac{\hbar}{\omega_i}} \sum_{\alpha} \frac{K_{\alpha i}}{\sqrt{M^{\alpha}}} d_{nn'}^{\alpha}(\bm{R}^\text{eq}) \label{eqn:lambda},
\end{gather}
where $\bm{R}^\text{eq}$ denotes the ground-state equilibrium geometry, $M^\alpha$ is the mass associated with the $\alpha$-th Cartesian coordinate, and the matrix $\bm{K}$ represents the normal modes expressed in mass-weighted coordinates. The quadratic coefficients $\gamma_i^{(n)}$, in contrast, were obtained by a least-squares fit to adiabatic potential energy data points computed with RI-CC2 as a function of the mode coordinate $Q_i$.

\subsection{\label{subsec:fssh}Fewest switches surface hopping}
In both the VC model and \emph{ab initio} on-the-fly calculations, the fewest-switches surface hopping (FSSH) algorithm\cite{tully_jcp1990_fssh} was used to propagate the nonadiabatic dynamics. The adiabatic electronic coefficients $c_n(t)$ were propagated according to the time-dependent Schr\"{o}dinger equation
\begin{equation}
    \im \hbar \dv{c_n}{t} = E_n(\bm{X}) c_n(t) - \im \hbar \sum_{n'\alpha} \dot{X}^{\alpha} d_{nn'}^{\alpha}(\bm{X}) c_{n'}(t), 
\end{equation}
where $\bm{X}$ and $\dot{\bm{X}}$ denote the nuclear coordinates and velocities, respectively; $E_n(\bm{X})$ is the adiabatic energy of state $n$; and $\bm{d}_{nn'}(\bm{X})$ represents the NACV between states $n$ and $n'$. 
The nuclear degrees of freedom were propagated using Newtonian equations of motion, with the force given by the gradient $-\partial E_a/\partial\bm{X}$ evaluated for the active state $a$. Decoherence effects were incorporated using the energy-based decoherence (EDC) correction of Granucci and Persico\cite{granucci_jcp2007_decoherence,granucci_jcp2010_decoherence} with the decoherence parameter $\alpha = 0.1$ Hartree. The hopping probability from the active state $a$ to another state $k$ is given by\cite{tully_jcp1990_fssh}
\begin{equation}
    \gamma_{k\gets a} = \max\left(2\sum_{\alpha} \dot{X}^{\alpha}\Re[d_{ak}^{\alpha}(\bm{X}) c_k(t) c_{a}^{*}(t)], 0\right) / \abs{c_{a}(t)}^2.
\end{equation}
When an attempted hop to state $k$ was successful, the velocity was rescaled along the direction of the NACV $\bm{d}_{ak}(\bm{X})$ to conserve total energy. If the hop was frustrated, the velocity component along $\bm{d}_{ak}(\bm{X})$ was reversed, following the procedure described in Ref.~\cite{schiffer_jcp1994_velocity_reverse}.

To extract the chemically intuitive diabatic states from adiabatic TSH simulations, we require a diabatic representation.  The transition-dipole-based scheme of Medders \emph{et al.} \cite{medders_jpca2017_diabatization} is well-suited to this system, given the distinct bright/dark characters of the three states. However, the procedure described in Ref.~\cite{medders_jpca2017_diabatization} cannot be directly applied in this work because RI-CC2 is a non-Hermitian theory in which the dipole moments are not uniquely defined (see Sec.~\ref{subsec:ricc2}). To address this issue, we propose an approximate approach that utilizes both the right and left transition dipole moments from RI-CC2 (Eq.~\ref{eqn:dipoles}). Specifically, we construct the diabatic states by diagonalizing a symmetrized transition dipole moment matrix. Following the prescription of Medders \emph{et al.}, we build the dot product matrix $D$ as
\begin{equation}
    D = \sum_{n} \sum_{n'} \frac{\bm{\mu}_{0\gets n} \cdot \bm{\mu}_{n'\gets0} + \bm{\mu}_{0\gets n'} \cdot \bm{\mu}_{n\gets0} }{2} \ketbra{\text{S}_n}{\text{S}_{n'}}.
\end{equation}
The eigenvectors of $D$ define the adiabatic-to-diabatic transformation. After appropriate reordering and transposition, we obtain the transformation matrix $U$ that converts between the adiabatic and diabatic representations. As demonstrated in Sec.~\ref{sec:sim_res}, this diabatization approach performs well for our system.

Finally, to compute diabatic populations from the adiabatic electronic coefficients, we used the third approach described in Ref.~\cite{landry_jcp2013_diabatic}:
\begin{equation}
    P_d =  \abs{U_{d a}}^2 + \sum_{n<n'} 2\Re[U_{dn} c_n c_{n'}^{*}U_{dn'}^{*}],
\end{equation}
where $P_d$ is the population of diabatic state $d$ and $a$ denotes the active adiabatic state.

\subsection{\label{subsec:dann}Diabatic Artificial Neural Network}
The computational cost of ab initio RI-CC2 TSH precludes the large trajectory ensembles required for statistically robust analysis of the coupled electronic--nuclear dynamics. We therefore accelerate the simulations using a diabatic artificial neural network (DANN) force field \cite{axelrod_natcomm2022_dann}, trained to reproduce the RI-CC2 diabatic Hamiltonian while enabling rapid evaluations of energies, forces, and NACVs. The DANN combines an equivariant graph neural network\cite{pmlr-v139-schutt21a} with a diabatic readout \cite{williams_jcp2018_diab_nn,guan_pccp2019_diab_nn,shu_jcp2020_diab_nn}. From molecular geometries, the model generates equivariant features up to three-body terms (distances and angles) within a cutoff radius $r_\text{cut}$. The network outputs a diabatic Hamiltonian matrix; diagonalization yields adiabatic energies $E_{n}(\bm{R})$, and automatic differentiation combined with the Hellmann--Feynman theorem provides forces $-\partial E_n / \partial \bm{R}$ and NACVs $\bm{d}_{mn}$.

The loss functions used in this work are $\mathcal{L}_\text{core}$, $\mathcal{L}_\text{diab}$, and $\mathcal{L}_\text{nacv}$. Here, $\mathcal{L}_\text{core}$ penalizes errors in the adiabatic energies and their gradients;  $\mathcal{L}_\text{diab}$ penalizes errors in the diagonal elements of the diabatic Hamiltonian; and $\mathcal{L}_\text{nacv}$ penalizes errors in the nonadiabatic coupling vectors. The explicit expressions for these loss functions are provided in Sec. S6 of the Supporting Information (SI). To robustly train these multiple objectives, we employed Jacobian descent \cite{quinton_arxiv_jacobian} of the multiple losses rather than optimizing a single scalar loss, as implemented in the torchjd package.

\section{\label{sec:comp_details}Computational details}
All electronic structure calculations---including ground-state energies \cite{zhao2024_sri_gs}, excitation energies\cite{zhao2024_sri_es}, transition dipole operators and oscillator strengths\cite{zhao2025_srifosc}, ground- and excited-state gradients and non-adiabatic coupling vectors \cite{zhao2025_srifosc,zhao2025_sri_grad_nac}---were performed using the Generalized Many-Body Perturbation Theory (GMBPT) module of Q-Chem~\cite{Epifanovsky2021_qchem5}. The basis set used in this work was cc-pVDZ~\cite{dunning_jcp1989_ccpvdz}. The geometric optimization and frequency analysis were performed with the GeomeTRIC package~\cite{wang_jcp2016_geometric}.

We first performed a ground-state geometry optimization to obtain the equilibrium reference geometry $\bm{R}^\text{eq}$ (see SI, Table S1). A subsequent frequency analysis at $\bm{R}^\text{eq}$ was then carried out to obtain the vibrational frequencies $\omega_i$ and the mass-weighted normal modes $\bm{K}$. Single-point calculations at the reference geometry provided the excitation energies $E_n^\text{eq}$, gradients $\eval{\partial E_n / \partial R_\alpha}_{\bm{R}=\bm{R}^\text{eq}}$, and nonadiabatic coupling vectors $\bm{d}_{nn'}(\bm{R}^\text{eq})$. These quantities, together with Eqs.~\ref{eqn:kappa} and~\ref{eqn:lambda}, yield the linear vibronic coupling constants $\kappa_i^{n}$ and $\lambda_i^{nn'}$. Based on the criterion $\kappa_i^{n}, \lambda_i^{nn'} > 0.001,\text{eV}$, we selected six tuning modes ($Q_{6a}$, $Q_{12}$, $Q_{1}$, $Q_{9a}$, $Q_{8a}$, $Q_{2}$) and seven coupling modes ($Q_{6b}$, $Q_{4}$, $Q_{10a}$, $Q_{5}$, $Q_{3}$, $Q_{8b}$, $Q_{7b}$). For these seven coupling modes and the $Q_{6a}$ tuning mode, 20 ab initio points were calculated over the displacement range $Q_i \in [-5,5]$. A least-squares fit to these points yielded the quadratic coupling coefficients $\gamma_i^{n}$ for these eight modes. The complete parameters are provided in Tables S2--S5 of the SI.

For the VC model, we performed exact quantum dynamics using the Matrix Product State Quantum Dynamics (MPSQD) package introduced in Ref.~\cite{guan_jcp2024_mpsqd}. The initial electronic state was prepared in the $B_{2u}$ diabatic state, corresponding to a vertical excitation scenario, and all harmonic oscillators were initialized in their ground vibrational state. A bond dimension of 120 and a harmonic oscillator basis size of 40 were required to converge the dynamics (SI Sec. S4, Figure S1). The time-dependent variational principle~\cite{lubich_2015_tdvp,guan_jcp2024_mpsqd} was used to propagate the wavefunction with a time step of 0.25 fs.

In addition to the exact quantum dynamics, we also performed trajectory surface hopping (TSH) simulations for the VC model using an in-house code. To ensure consistency with the quantum dynamics calculations, initial geometries and velocities of the normal modes were sampled from a Wigner distribution of the ground state. The electronic initial condition was prepared in the $B_{2u}$ diabatic state and transformed to the adiabatic representation using the transformation matrix $U$. The initial active state was then randomly sampled from the resulting adiabatic wavefunction. Nuclear positions and velocities were propagated using the velocity-Verlet algorithm with a time step of 2 atomic units (ca. 0.048 fs), and a swarm of 1000 trajectories was simulated for 200 fs.

To perform ab initio TSH simulations for pyrazine, we interfaced the GMBPT code with the SHARC package.\cite{mai2018NonadiabaticDynamicsSHARC,sharc4code2025} The three lowest excited adiabatic states, $\text{S}_1$, $\text{S}_2$, and $\text{S}_3$, were included in our simulations. Initial geometries and velocities were sampled from a Wigner distribution of the harmonic normal modes at 300 K, based on the ground-state frequency analysis. The initial electronic coefficients and active state were obtained using the same procedure as for the VC model TSH simulations. Nuclear propagation employed the velocity-Verlet algorithm with a time step of 0.5 fs. 

A total of 100 \emph{ab initio} trajectories were simulated for 100 fs, of which 98 completed the full time window. Two trajectories ended early, at $t = 63.0$ and $72.5$ fs, because the RI-CC2 eigensolver failed to converge near degenerate $\text{S}_2/\text{S}_3$ geometries. We confirmed that these two cases involve complex-conjugate root pairs---exactly the defective region where our real-arithmetic RI-CC2 implementation cannot converge reliably. We refer the interested reader to SI Sec. S14 for a detailed analysis. Both trajectories were therefore excluded from all statistical analyses. For the pyrazine TSH simulations with a fixed time step of 0.5 fs, such eigensolver failures are rare (2 of 19,971 geometries). Nonetheless, users of RI-CC2 should be aware of this potential instability when applying the method to other systems.

A DANN model was trained to perform TSH simulations with RI-CC2 accuracy with much reduced computational cost. Initially, the 500 geometries sampled from the Wigner distribution were used to train a primitive model. This model was then iteratively improved via an active learning procedure following Ref.~\cite{zhang_prm2019_dpgen} until the internal conversion dynamics converged. The converged dataset consists of 2545 geometries, randomly split into a training and validation sets of 2288 and 257 geometries, respectively. Detailed training procedures are provided in Sec. S6  of the SI. The performance of the DANN model for the validation set was shown in Figure S3 of the SI. This model was employed to perform surface hopping simulations of 2000 trajectories for 200 fs with a time step of 0.25 fs, where the simulation procedures are identical to those of the \emph{ab initio} simulations.

\section{\label{sec:sim_res}Results and discussion}

\subsection{\label{subsec:pes_abs}\emph{Ab initio} potential energy surfaces and absorption spectrum}
The role of vibrational motion in the ultrafast internal conversion of pyrazine is a fundamental problem that remains under debate. Two open questions persist: (1) whether the dark $A_\text{1u}$ state plays a significant role, and (2) which vibrational modes contribute to the quantum beats observed in experiments.

\begin{figure*}[htbp]
    \centering
    \includegraphics[width=0.65\linewidth]{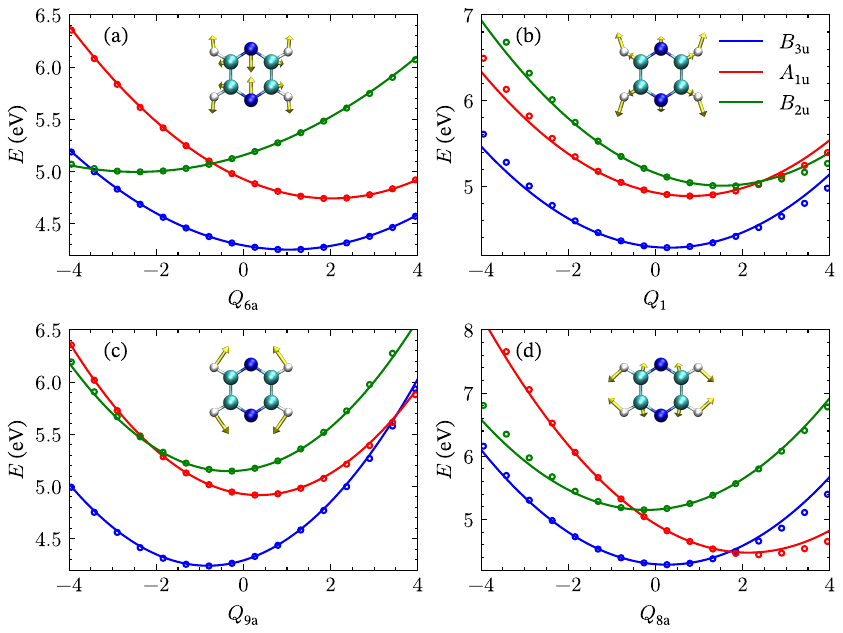}
    \caption{Potential energy surfaces along the four most significant tuning modes: $Q_{\text{6a}}$ (a), $Q_{\text{1}}$ (b), $Q_{\text{9a}}$ (c), and $Q_{\text{8a}}$ (d). The green, red, and blue colors represent the diabatic states $B_{\text{3u}}$, $A_{\text{1u}}$, and $B_{\text{2u}}$, respectively. Open circles denote the \emph{ab initio} data points, while solid lines correspond to the VC model.}
    \label{fig:pes}
\end{figure*}

To address these questions, we first plot one-dimensional potential energy surfaces (PES) cut along the four fully symmetric vibrational tuning modes in Figure~\ref{fig:pes}. These potential energies and their implications for the dynamics have been comprehensively discussed by Sala \emph{et al.} using extended multi-configuration quasi-degenerate second-order perturbation theory (XMCQDPT2)~\cite{sala_pccp2014_RoleLowlyingDark} and by Xie \emph{et al.} using the algebraic diagrammatic construction ADC(2)~\cite{xie2019_assessing}. Here, we complement their results with RI-CC2 calculations. A direct comparison of the RI-CC2 potential energy surfaces with the XMCQDPT2 reference is provided in the SI (Figure S13), confirming consistent state ordering along all four tuning modes between these two methods. We note, however, that this comparison establishes consistency among theoretical methods rather than an absolute determination of the correct state ordering; resolving the true state ordering would require further investigation beyond the scope of the present work\cite{state_ordering2026}.

\begin{figure}[htbp]
    \centering
    \includegraphics[width=0.75\linewidth]{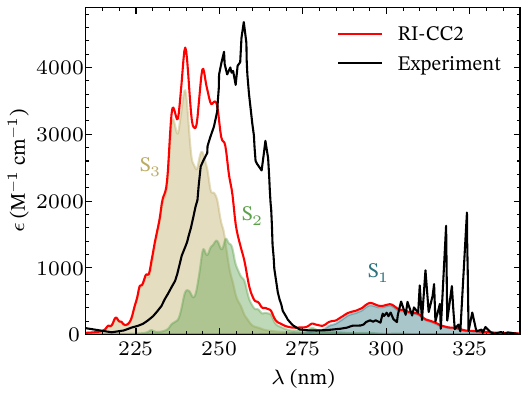}
    \caption{Absorption spectrum of pyrazine. The solid red line shows the calculated spectrum (RI-CC2), and the solid black line shows the experimental spectrum. The RI-CC2 result is further decomposed into the contributions from the $\text{S}_1$, $\text{S}_2$, and $\text{S}_3$ states, shown as filled areas. A Lorentzian broadening parameter $\Gamma= 0.04\,\text{eV}$ was applied. Experimental absorption spectrum based on numerical data from Ref.~\cite{keller-rudek2013MPIMainzUVVIS}, distributed under the Creative Commons Attribution 3.0 (CC BY 3.0) License. The original experimental data were reported in Ref.~\cite{bolovinos_jmolspec1984_uvvis}.
}
    \label{fig:abs_spectra}
\end{figure}

As pointed out by Xie \emph{et al.},\cite{xie2019_assessing} RI-CC2 overestimates the vertical excitation energies to the $B_\text{3u}$ and $B_\text{2u}$ states by ca. 0.3 eV (see Table S2 in the SI). This overestimation is further evident in the ab initio absorption spectrum shown in Figure~\ref{fig:abs_spectra}. When static disorder is accounted for through sampling over 500 configurations from a Wigner distribution, RI-CC2 faithfully reproduces the absolute molar absorptivities and the bandwidths of the two bright states, albeit with a consistent blueshift of ca. 0.3 eV.

RI-CC2 predicts that the $B_\text{2u}/A_\text{1u}$ conical intersections (CIs) are located much closer to the Franck--Condon geometry than the $B_\text{2u}/B_\text{3u}$ crossings (Figure~\ref{fig:pes}), facilitating population transfer to the $A_\text{1u}$ state after vertical excitation. However, the coupling strength between the two bright states $B_\text{2u}$ and $B_\text{3u}$ via mode $Q_\text{10a}$ ($\lambda_{10a}=0.28$ eV) is significantly larger than the coupling between $B_\text{2u}$ and $A_\text{1u}$ via modes $Q_4$ and $Q_5$ ($\lambda_4 = 0.0668$ eV, $\lambda_5 = 0.0052$ eV). This leads to competition between population of the $A_\text{1u}$ and $B_\text{3u}$ states after the initial excitation. Moreover, the PES along the tuning mode $Q_\text{8a}$ features a CI between the $A_\text{1u}$ and $B_\text{3u}$ states, and these two states are strongly coupled via mode $Q_\text{8b}$ with a significant coupling constant of $\lambda_{8b} = -0.2329$ eV. This indicates that once the dark $A_\text{1u}$ state is populated, population will oscillate between the $A_\text{1u}$ and $B_\text{3u}$ states. These RI-CC2 predictions are consistent with previous XMCQDPT2~\cite{sala_pccp2014_RoleLowlyingDark} and ADC(2)~\cite{xie2019_assessing} studies.

In summary, the static calculations with RI-CC2, together with the dynamics presented in Secs.~\ref{subsec:vc_dynamics} and~\ref{subsec:on_the_fly}, support that the dark $A_\text{1u}$ state actively participates in the internal conversion of pyrazine.

\subsection{\label{subsec:vc_dynamics}Vibronic coupling model dynamics}
\begin{figure}[htbp]
    \centering
    \includegraphics[width=0.85\linewidth]{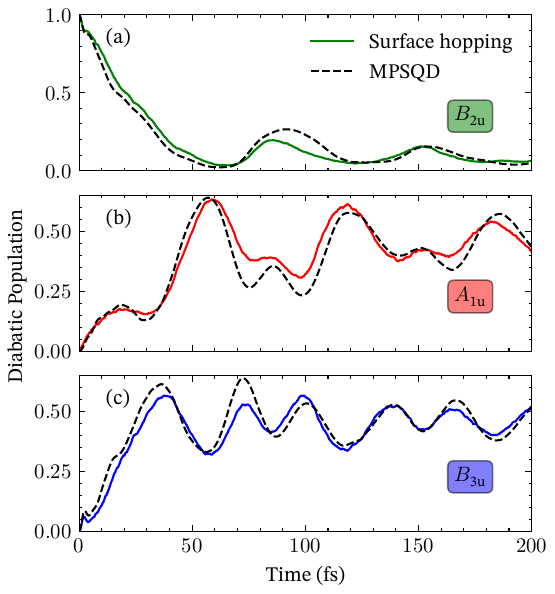}
    \caption{Time evolution of the diabatic populations of the $B_\text{2u}$ (a), $A_\text{1u}$ (b), and $B_\text{3u}$ (c) states. Colored solid lines represent the surface hopping results, while black dashed lines correspond to the exact MPSQD results.}
    \label{fig:vc_diab_pop}
\end{figure}

Figure~\ref{fig:vc_diab_pop} shows the diabatic populations of the VC model computed with exact MPSQD and TSH. TSH captures all the essential features of the quantum dynamics at a significantly reduced computational cost, justifying its use in describing the ultrafast internal conversion of pyrazine.

Following vertical excitation, both the $B_\text{3u}$ and $A_\text{1u}$ states become populated within the first 10 fs, consistent with the PES landscape shown in Figure~\ref{fig:pes}. However, due to the stronger coupling between $B_\text{2u}$ and $B_\text{3u}$, the $B_\text{3u}$ state becomes dominant by approximately 40 fs. After this point, the population of the bright $B_\text{2u}$ state is nearly depleted, while the $A_\text{1u}$ and $B_\text{3u}$ states become comparably populated, and an oscillation between these two states persists up to 200 fs. Moreover, the dynamics reproduce the two-component beating reported for the pyrazine VC model~\cite{tsuru_jcp2019_nexafs}: a dominant oscillation with a period of $\approx$60 fs that was assigned to Q$_\text{6a}$ and a minor $\approx$30 fs component assigned to Q$_\text{9a}$.

Notably, Figure S2 of the SI shows that the $Q_\text{8a}$ and $Q_\text{9a}$ oscillations are phase-locked with the $A_\text{1u}$ and $B_\text{3u}$ populations in the $t > 40$ fs window. This behavior was reported for the $Q_\text{8a}$ mode in the XMCQDPT2 VC model \cite{sala_pccp2014_RoleLowlyingDark,xie2019_assessing}. Such correlated, phase-locked motion of the electronic and nuclear degrees of freedom observed here---the oscillation of the $A_\text{1u}$ and $B_\text{3u}$ populations together with the ensemble-averaged $Q_\text{8a}$ and $Q_\text{9a}$ coordinates---constitutes a classical vibronic coherence in the sense of Miller~\cite{miller_jcp2012_coherence}. It is distinct from quantum coherence arising from the interference of nuclear wave-packet amplitudes, which lies beyond the classical-nuclei description used here. Finally, we note that the exact quantum dynamics (dashed lines in Figure~S2) exhibit the same phase locking, indicating that this correlated classical motion is the classical limit of a genuinely quantum coherence.

\subsection{\label{subsec:on_the_fly}\emph{Ab initio} surface hopping dynamics}

\begin{figure}[htbp]
    \centering
    \includegraphics[width=0.85\linewidth]{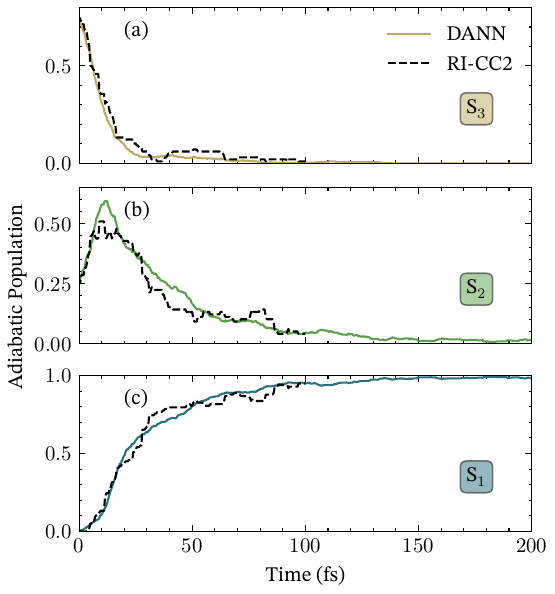}
    \caption{Time evolution of the adiabatic populations of the $\text{S}_3$ (a), $\text{S}_2$ (b), and $\text{S}_1$ (c) states. Colored solid lines represent the surface hopping results from the DANN model, while black dashed lines correspond to the results from RI-CC2.}
    \label{fig:on_the_fly_adiab_pop}
\end{figure}

Having validated the mechanistic picture and the TSH implementation against exact dynamics in the reduced space (Sec.~\ref{subsec:vc_dynamics}), we now test whether the correlated population dynamics survive in full dimensionality. To make this computationally feasible, we use the DANN-accelerated model verified below. 
 
\begin{figure}[htbp]
    \centering
    \includegraphics[width=0.85\linewidth]{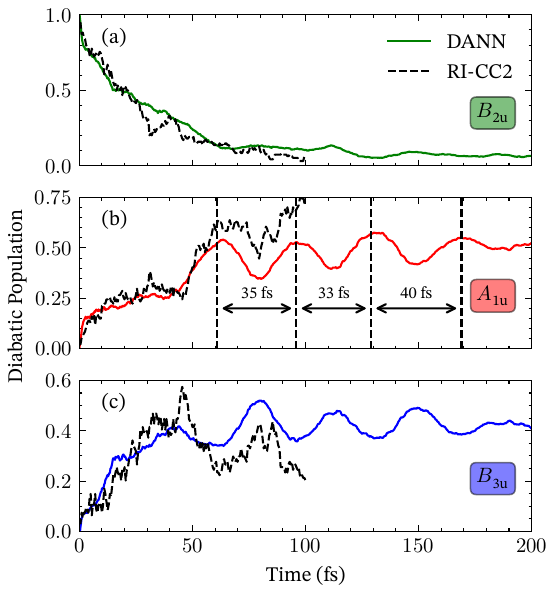}
    \caption{Time evolution of the diabatic populations of the $B_\text{2u}$ (a), $A_\text{1u}$ (b), and $B_\text{3u}$ (c) states.  Colored solid lines represent the surface hopping results from the DANN model, while black dashed lines correspond to the results from RI-CC2.}
    \label{fig:on_the_fly_diab_pop}
\end{figure}

Figures~\ref{fig:on_the_fly_adiab_pop} and \ref{fig:on_the_fly_diab_pop} show the time evolution of the adiabatic and diabatic state populations; the DANN results are in excellent agreement with the reference RI-CC2 calculations. Moreover, with the highly efficient DANN model, we simulated up to 2000 trajectories. Using a bootstrap resampling analysis, we confirmed that 500 trajectories yield smooth and statistically converged population dynamics (Figure S7). Bootstrap confidence intervals for the \emph{ab initio} ensemble-averaged populations, their convergence with the number of trajectories, and a decomposition of the statistical noise into finite-sampling and diabatization contributions are provided in Sec.~S9 of the SI.

The initial state was prepared as the bright $B_\text{2u}$ diabatic state following the procedures described in Secs.~\ref{subsec:fssh} and \ref{sec:comp_details}. This diabatic initialization results in an initial population distribution of approximately 75\% in $\text{S}_3$  and 25\% in $\text{S}_2$. The population of $\text{S}_3$ fully relaxes within the first $\approx$30 fs, while the $\text{S}_2$ population increases slightly over the first 10 fs before decaying rapidly within 100 fs. Consequently, the lowest excited state  $\text{S}_1$ becomes essentially fully populated after 100 fs, in accordance with Kasha’s rule~\cite{kasha_1950_kashasrule}.

\begin{figure*}[htbp]
    \centering
    \includegraphics[width=0.95\linewidth]{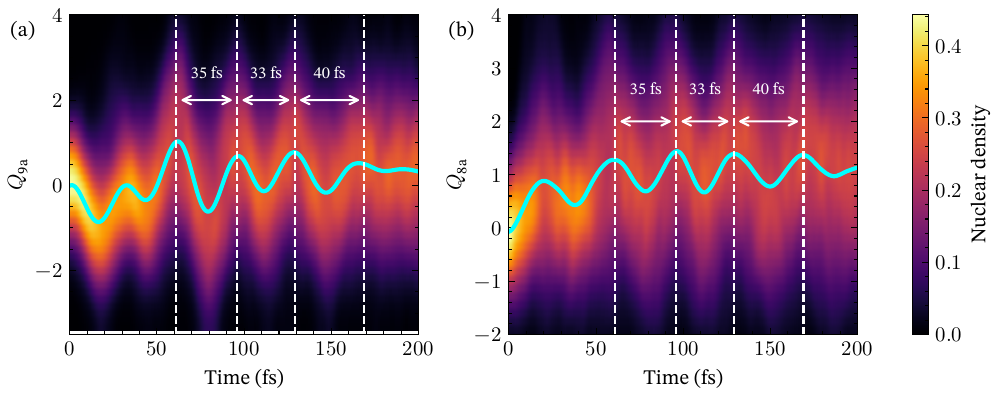}
    \caption{Time evolution of the nuclear density projected onto the $Q_\text{9a}$ (a) and $Q_\text{8a}$ (b) modes from the DANN simulations. The color map represents the nuclear density, while the cyan line shows the average value of the $Q_\text{9a}$ and $Q_\text{8a}$ coordinates as a function of time. The corresponding plot obtained with RI-CC2 is provided in Figure S5 of the SI.}
    \label{fig:heatmap_nuc}
\end{figure*}

Figure~\ref{fig:on_the_fly_diab_pop} shows the diabatic populations as a function of time from the ab initio on-the-fly dynamics. The overall behavior is similar to that of the VC model (Figure~\ref{fig:vc_diab_pop}), but with notable differences. Recurrences of the $B_\text{2u}$ population observed near 90 and 150 fs in the VC model are significantly suppressed in the full-dimensional model. Consequently, the $B_\text{2u}$ population decays exponentially with a time constant of 26 fs (see Figure S4 in the SI), in good agreement with the experimental value of $22 \pm 3$ fs reported by Horio \emph{et al.}~\cite{horio2009trpei,suzuki_jcp2010_trpei}.

The populations of the A$_\text{1u}$ and B$_\text{3u}$ states increase within the first 40 fs, after which they begin to oscillate. The A$_\text{1u}$ population in the on-the-fly results exhibits a more regular oscillation pattern than the alternating strong--weak oscillation observed in the VC model dynamics (Figure~\ref{fig:vc_diab_pop}). Specifically, the oscillation in the A$_\text{1u}$ state shows recurrence intervals of approximately 35,
33, and 40 fs. Notably, the dominant $\approx$60 fs oscillation characteristic of the reduced-dimensional VC model~\cite{tsuru_jcp2019_nexafs} is not apparent in the full-dimensional dynamics. This oscillation pattern in the A$_\text{1u}$ population correlates well with the dynamics of the Q$_\text{9a}$ and Q$_\text{8a}$ modes (Figure~\ref{fig:heatmap_nuc}), which tune the $A_\text{1u}$/$B_\text{3u}$ characters as they oscillate.

To determine the nature of this correlated electron--nuclear dynamics, we performed a direct analysis of the hopping events (see Sec.~S13 of the SI). This analysis shows that most hopping events occur within the first $\approx$60~fs. After this window, the population of the $S_1$ adiabatic surface quickly plateaus at 0.9, and hopping events rarely occur. Therefore, the sustained $A_\text{1u}$/$B_\text{3u}$ population oscillations originate from the oscillating diabatic character on the $S_1$ surface as the nuclear ensemble moves along the tuning modes. Specifically, the oscillatory motion of the Q$_\text{8a}$ and Q$_\text{9a}$ tuning modes modulates the observed population beats.

The assignment of the population beats to specific vibrational modes, however, is not unique across the literature. Karashima \emph{et al.} observed oscillations at 581 and 988 $\text{cm}^{-1}$ in their TRPES signal and assigned them to the $Q_\text{6a}$ and $Q_\text{1}$ modes, respectively~\cite{karashima_jacs2024_vibmotion}. Gelin \emph{et al.} found that the simulated ground-state bleaching signal exhibits oscillations with a period
of 33 fs, which they related to the $Q_\text{1}$ mode~\cite{gelinInitioSurfaceHoppingSimulation2021,liFinetuningLaserpulseFrequencies2025}. Xie \emph{et al.} associated the $A_\text{1u}$/$B_\text{3u}$ population oscillations with the $Q_\text{8a}$ mode~\cite{xie2019_assessing}. Tsuru \emph{et al.} assigned a dominant $\approx$60 fs oscillation to $Q_\text{6a}$ with a minor $\approx$30 fs component assigned to $Q_\text{9a}$~\cite{tsuru_jcp2019_nexafs}.

We note that the population oscillation arises from electrons and nuclei moving together on several coupled excited-state potential-energy surfaces. Therefore, ground-state frequencies alone cannot be used to assign oscillation patterns in the population. Figure~\ref{fig:heatmap_nuc} makes this point clear. Although the $Q_\text{9a}$ and $Q_\text{8a}$ modes have different ground-state frequencies, they oscillate in phase with the $A_\text{1u}$ and
$B_\text{3u}$ populations. Although $Q_\text{1}$ has a ground-state frequency (0.1265 eV, period 32.7 fs) corresponding to the $\approx$33~fs beating of the $A_\text{1u}/B_\text{3u}$ populations, Figure S6 of the SI shows that its nuclear density does not follow the population dynamics.

A direct comparison between the VC model and the on-the-fly simulations is provided in the SI (Figures S10--S12). For $Q_\text{8a}$, both the average displacement and the spread of the nuclear density distribution are comparable between the VC model and the on-the-fly dynamics, with the VC model showing a slightly earlier onset of the oscillations. For $Q_\text{9a}$, the time evolution of the average displacement differs between the two descriptions, with the VC model predicting a more contracted distribution than the on-the-fly dynamics. Together with the absence of the two-component beating pattern in the on-the-fly dynamics, this suggests that the reduced VC model may not fully represent the tuning contribution of $Q_\text{9a}$ to the nonadiabatic dynamics of pyrazine.

In summary, the \emph{ab initio} dynamics at the RI-CC2 level support the active role of the dark $A_\text{1u}$ state in the internal conversion of pyrazine. The $Q_\text{8a}$ mode is indeed correlated with the population beat between the $A_\text{1u}$ and $B_\text{3u}$ states. The RI-CC2 calculations further show that the $Q_\text{9a}$ mode is involved not only in the initial $B_\text{2u} \to B_\text{3u}$ population transfer, but also in the correlated dynamics. Notably, the two-component beating structure of the reduced-dimensional VC model is not present in full-dimensional simulations. The present on-the-fly results are in good agreement with the ADC(2) studies by Xie \emph{et al.}~\cite{xie2019_assessing} and Chang \emph{et al.}\cite{Chang2025}, despite both employing an overlap-based method for nonadiabatic dynamics rather than the NACV-based approach used here. A direct trajectory-level comparison of the ensemble-averaged adiabatic and diabatic populations with the published ADC(2) dynamics~\cite{xie2019_assessing} is provided in Sec.~S12 of the SI (Figure~S14), showing semi-quantitative agreement between the two methods.

\section{\label{sec:conclusion}Conclusion}

In this work, we investigated the ultrafast internal conversion dynamics of pyrazine using the RI-CC2/cc-pVDZ level of theory, enabled by our implementation of analytical gradients and nonadiabatic coupling vectors in Q-Chem.\cite{zhao2024_sri_es,zhao2024_sri_gs,zhao2025_srifosc,zhao2025_sri_grad_nac} Using RI-CC2, we constructed a low-dimensional VC model and performed \emph{ab initio} on-the-fly TSH dynamics, with the latter accelerated by a DANN machine-learning model. Both the VC model and the full-dimensional dynamics show that the dark $A_\text{1u}$ state actively participates in the internal conversion of pyrazine. Furthermore, RI-CC2 reveals that the population oscillations between the $A_\text{1u}$ and $B_{3u}$ states are correlated with the motion of the $Q_\text{8a}$ and $Q_\text{9a}$ modes; the $Q_\text{9a}$ dynamics, however, differ between the  VC and on-the-fly descriptions.

Looking forward, the high-quality dataset of energies, forces, and NACVs generated here should prove valuable for developing machine-learning models for excited states. Light--matter interactions under one-\cite{wangNonadiabaticDynamicsMetal2023a,wangNonadiabaticDynamicsMetal2023,mosallanejadTwomodeFloquetRedfieldQuantum2025} or two-frequency\cite{mosallanejadTwomodeFloquetRedfieldQuantum2025,hanTwoModeFloquetFewest2026} periodic driving can be studied with our recently developed Floquet surface hopping method. Moreover, the stochastic variant sRI-CC2 reduces the computational scaling from $\order{N^5}$ to $\order{N^3}$, suggesting that excited-state dynamics could be extended to much larger systems with appropriate parallelization. Further developments along these lines are ongoing.

\section*{Author contributions}
Conceptualization: R.-H., W. D.; Data curation: R.-H., R. S.; Formal Analysis: R.-H.; Funding acquisition: W. D.; Software: R.-H., C. Z., R. S.; Supervision: W. D.; Writing -- original draft: R.-H., W. D.; Writing -- review \& editing: R.-H., C. Z., W. D.

\section*{Conflicts of interest}
The authors declare no conflicts of interest regarding this manuscript.

\section*{Data availability}
The SI contains quantum chemistry data for the excited states of pyrazine at the RI-CC2/cc-pVDZ level of theory, details of the machine learning model and training procedure, and additional figures. All data required to reproduce the figures in this work, the ab initio dataset for the ultrafast internal conversion of pyrazine (singlet-state energies, gradients, and interstate NACVs), as well as the trained DANN model are publicly available on the Figshare repository at:
\href{https://doi.org/10.6084/m9.figshare.31866094}{DOI: 10.6084/m9.figshare.31866094}. The repository also contains the raw surface-hopping trajectory histories (including the potential energies at every step), the input and output files of the two failing RI-CC2 calculations together with corresponding Turbomole calculations at both failing geometries, the per-step classification files of the eigensolver stability audit (Sec.~S14 of the SI), and the analysis scripts. Other data are available from the corresponding author upon reasonable request.

\begin{acknowledgement}
We are grateful to Graham Worth for the discussion of quantum dynamics and to Weiwei Xie for valuable discussions on the calculation of diabatic populations. W.D. acknowledges the support from National Natural Science Foundation of China (No. 22361142829 and No. 22273075) and Zhejiang Provincial Natural Science Foundation (No. XHD24B0301). We thank Westlake University Supercomputer Center for the facility support and technical assistance.
\end{acknowledgement}

\bibliography{ref}

\end{document}


\maketitle

\clearpage
\tableofcontents
\clearpage

\section{Quantum chemistry data for pyrazine}
\begin{table}[htbp]
\caption{\label{stab:pyrazine_gs_xyz} Optimized Cartesian coordinates (in Å) of pyrazine ground state equilibrium geometry calculated using RI-CC2/cc-pVDZ.}
\begin{tabular}{lrrr}
    \toprule
    H &  2.08008529  &  0.00083553 & -1.26063437 \\
    H & -2.08008533  & -0.00083552 & -1.26063432 \\
    H &  2.08008529  &  0.00083553 &  1.26063437 \\
    H & -2.08008533  & -0.00083552 &  1.26063432 \\
    C &  1.13699090  &  0.00045672 & -0.70247262 \\
    C & -1.13699089  & -0.00045672 & -0.70247264 \\
    C &  1.13699090  &  0.00045672 &  0.70247262 \\
    C & -1.13699089  & -0.00045672 &  0.70247264 \\
    N & -0.00000001  & -0.00000000 & -1.43242914 \\
    N & -0.00000001  & -0.00000000 &  1.43242914 \\
    \bottomrule
\end{tabular}
\end{table}

\clearpage


\begin{table}[htbp]
\caption{\label{stab:gs_excitation} Excitation energies and oscillator strengths of pyrazine calculated at ground state equilibrium geometry using RI-CC2/cc-pVDZ.}
\begin{threeparttable}
\begin{tabular}{cccccc}
\toprule
   & \multicolumn{2}{c}{Excitation energy (eV)} & \multicolumn{2}{c}{Oscillator strength} \\
   \cmidrule(lr){2-3}
   \cmidrule(lr){4-5}
   & RI-CC2 & Experiment         & RI-CC2  & Experiment               \\ \midrule
$\text{S}_1$ ($B_\text{3u})$ & 4.293  & 3.97\tnote{$\alpha$} & 0.0055  & 0.006\tnote{$\gamma$}  \\
$\text{S}_2$ ($A_\text{1u})$ & 4.928  & ...                  & 0.0000  & ...                    \\
$\text{S}_3$ ($B_\text{2u})$ & 5.156  & 4.81\tnote{$\beta$}  & 0.0711  & 0.062\tnote{$\gamma$}  \\ \bottomrule
\end{tabular}
\begin{tablenotes}
    \item[$\alpha$] 0-0 transition of 3.83 \text{eV} (Ref.~\cite{bolovinos_jmolspec1984_uvvis}) plus the average reorganization energy from Ref.~\cite{weberInitioDensityFunctional1999}.
    \item[$\beta$] Vertical excitation energy from Ref.~\cite{bolovinos_jmolspec1984_uvvis}.
    \item[$\gamma$] Experiment oscillator strength from Ref.~\cite{bolovinos_jmolspec1984_uvvis}.
\end{tablenotes}
\end{threeparttable}
\end{table}

\clearpage

\begin{table}[htbp]
\caption{\label{stab:freq} Calculated (RI-CC2/cc-pVDZ) and experimental (Ref.~\cite{yamazaki1983IntramolecularElectronicRelaxation}) vibrational frequencies in $\text{eV}$ for the ground state of pyrazine.}
\begin{tabular}{ccrr}
\toprule
       Symmetry &                 Mode & Experiment &     RI-CC2 \\
\midrule
        $A_{g}$ &    $\nu_{\text{6a}}$ &     0.0739 &     0.0734 \\
                &     $\nu_{\text{1}}$ &     0.1258 &     0.1265 \\
                &    $\nu_{\text{9a}}$ &     0.1525 &     0.1544 \\
                &    $\nu_{\text{8a}}$ &     0.1961 &     0.1995 \\
                &     $\nu_{\text{2}}$ &     0.3788 &     0.3999 \\
       $B_{1g}$ &   $\nu_{\text{10a}}$ &     0.1139 &     0.1167 \\
       $B_{2g}$ &     $\nu_{\text{4}}$ &     0.0937 &     0.0944 \\
                &     $\nu_{\text{5}}$ &     0.1219 &     0.1207 \\
       $B_{3g}$ &    $\nu_{\text{6b}}$ &     0.0873 &     0.0870 \\
                &     $\nu_{\text{3}}$ &     0.1669 &     0.1679 \\
                &    $\nu_{\text{8b}}$ &     0.1891 &     0.1923 \\
                &    $\nu_{\text{7b}}$ &     0.3769 &     0.3974 \\
        $A_{u}$ &   $\nu_{\text{16a}}$ &     0.0423 &     0.0403 \\
                &   $\nu_{\text{17a}}$ &     0.1190 &     0.1219 \\
       $B_{1u}$ &    $\nu_{\text{12}}$ &     0.1266 &     0.1340 \\
                &   $\nu_{\text{18a}}$ &     0.1408 &     0.1639 \\
                &   $\nu_{\text{19a}}$ &     0.1840 &     0.1777 \\
                &    $\nu_{\text{13}}$ &     0.3734 &     0.3994 \\
       $B_{2u}$ &   $\nu_{\text{18b}}$ &     0.1318 &     0.1264 \\
                &    $\nu_{\text{14}}$ &     0.1425 &     0.1426 \\
                &   $\nu_{\text{19b}}$ &     0.1756 &     0.1851 \\
                &   $\nu_{\text{20b}}$ &     0.3798 &     0.3973 \\
       $B_{3u}$ &   $\nu_{\text{16b}}$ &     0.0521 &     0.0513 \\
                &    $\nu_{\text{11}}$ &     0.0973 &     0.0990 \\
\bottomrule
\end{tabular}
\end{table}

\clearpage

\section{\emph{Ab initio} absorption spectra of pyrazine}
The absorption spectrum was averaged over $N_\text{samp}=500$ configurations sampled from a Wigner distribution. It was computed using the following formula:
\begin{equation}
\epsilon(\omega) = \frac{1}{N_\text{samp}}\sum_{i}^{N_\text{samp}} \sum_{n=1}^{3} \frac{f_{n\gets0}^{i}}{C\pi}\frac{\Gamma/2}{(\omega-\omega_{ni})^2+(\Gamma/2)^2},
\end{equation}
where $f_{n\gets0}^{i}$ and $\omega_{ni}$ are, respectively, the oscillator strength and excitation energy for the $\text{S}_n$ state of the $i$-th sample; $\Gamma$ is the full width at half maximum (FWHM). The constant $C = 3.483\times10^{-5}\ \text{M}\cdot\text{cm}\cdot\text{eV}^{-1}$ converts the result to molar absorptivity units.~\cite{bolovinos_jmolspec1984_uvvis,keller-rudek2013MPIMainzUVVIS}

\clearpage

\section{Vibronic coupling model parameters for Pyrazine}

\begin{table}[htbp]
    \centering
    \caption{\label{stab:diag_couplings} Diagonal vibronic coupling coefficients (in $\text{eV}$) at the RI-CC2/cc-pVDZ level of theory.}
\begin{tabular}{rrrrrrr}
\toprule
     & \multicolumn{2}{c}{$B_{3u}$} & \multicolumn{2}{c}{$A_{1u}$} & \multicolumn{2}{c}{$B_{2u}$}  \\
     \cmidrule(lr){2-3}
     \cmidrule(lr){4-5}
     \cmidrule(lr){6-7} 
     & \multicolumn{1}{c}{$\kappa_i^0$}  
     & \multicolumn{1}{c}{$\gamma_i^0$} 
     & \multicolumn{1}{c}{$\kappa_i^1$} 
     & \multicolumn{1}{c}{$\gamma_i^1$} 
     & \multicolumn{1}{c}{$\kappa_i^2$}  
     & \multicolumn{1}{c}{$\gamma_i^2$} \\
\midrule 
$\nu_{6a}$  & 0.0781 & 0.0012  & 0.1841  & 0.0090  & -0.1305  & -0.0100\\
$\nu_{6b}$  & -      & -0.0079 & -       & -0.0167 & -        & -0.0051\\
$\nu_{4}$   & -      & -0.0267 & -       & -0.0253 & -        & -0.0384\\
$\nu_{10a}$ & -      & -0.0065 & -       & -0.0346 & -        & -0.0023\\
$\nu_{5}$   & -      & -0.0076 & -       & -0.0266 & -        & -0.0129\\
$\nu_{12}$  & -      & -       & -0.0011 & -       & -0.0022  & -      \\
$\nu_1$     & 0.0405 & -       & 0.0998  & -       & 0.1927   & -      \\
$\nu_{9a}$  & -0.1255& -       & 0.0551  & -       & -0.0514  & -      \\
$\nu_3$     & -      & -0.0038 & -       & -0.0036 & -        & 0.0025 \\
$\nu_{8b}$  & -      & -0.0019 & -       & -0.0198 & -        & 0.0022 \\
$\nu_{8a}$  & -0.0535& -       & -0.4246 & -       & 0.0424   & -      \\
$\nu_{7b}$  & -      & 0.0173  & -       & 0.0165  & -        & 0.0173 \\
$\nu_2$     & -0.0198& -       & -0.0711 & -       & -0.0159  & -      \\
\bottomrule
\end{tabular}
\end{table}

\clearpage

\begin{table}[htbp]
    \centering
    \caption{\label{stab:offdiag_couplings} Off-diagonal linear vibronic coupling coefficients $\lambda_i^{nn'}$ (in $\text{eV}$) at the RI-CC2/cc-pVDZ level of theory.}
\begin{tabular}{rrrr}
\toprule
     & \multicolumn{1}{c}{$B_{3u}/A_{1u}$} 
     & \multicolumn{1}{c}{$B_{3u}/B_{2u}$} 
     & \multicolumn{1}{c}{$A_{1u}/B_{2u}$} \\
     \midrule
$\nu_{6b}$  & 0.0075         & -               & -              \\
$\nu_4$     & -              & -               & 0.0668         \\
$\nu_{10a}$ & -              & 0.1846          & -              \\
$\nu_5$     & -              & -               & 0.0052         \\
$\nu_3$     & -0.0702        & -               & -              \\
$\nu_{8b}$  & -0.2329        & -               & -              \\
$\nu_{7b}$  & -0.0297        & -               & -              \\
\bottomrule
\end{tabular}
\end{table}

\clearpage


\section{Quantum dynamics of vibronic coupling model}

\begin{figure}[htbp]
    \centering
    \includegraphics[width=0.75\linewidth]{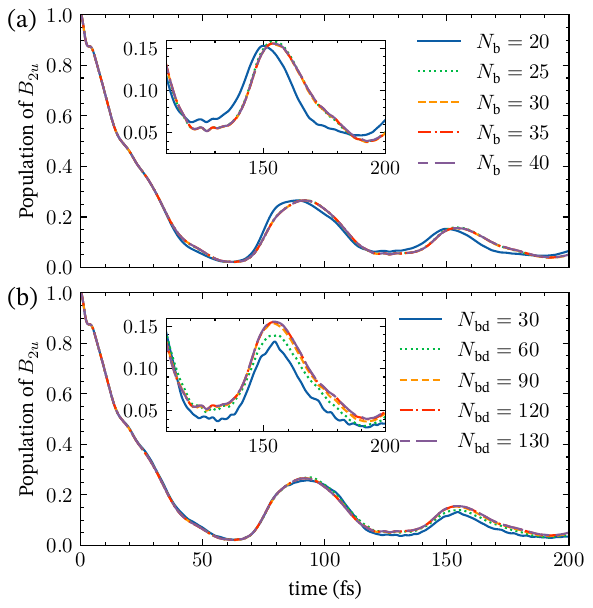}
    \caption{Convergence of the MPSQD dynamics with respect to the Hermite basis size $N_\text{b}$ and the bond dimension $N_\text{bd}$. Panel (a) shows the convergence with respect to $N_\text{b}$ for a fixed bond dimension of $N_\text{bd}=120$. Panel (b) shows the convergence with respect to $N_\text{bd}$ for a fixed Hermite basis size of $N_\text{b}=40$.}
    \label{sfig:mpsqd_convergence}
\end{figure}

\clearpage

\section{The nuclear dynamics of the vibronic coupling model}
\begin{figure}
    \centering
    \includegraphics[width=0.75\linewidth]{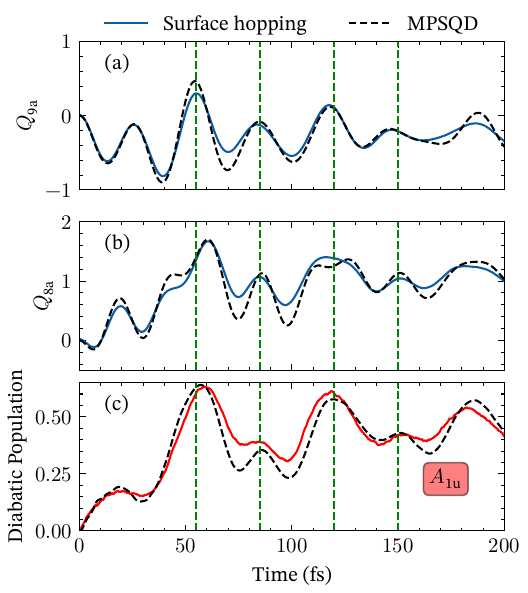}
    \caption{Time evolution of the expectation values $\expval{Q_\text{9a}}$ (a) and $\expval{Q_\text{8a}}$ (b), together with the diabatic population of the $A_\text{1u}$ state (c). Solid lines represent the surface hopping results, while dashed lines represent the exact MPSQD results.}
    \label{sfig:nuclear_vc}
\end{figure}

\clearpage

\section{Loss functions of the DANN model and training}
The explicit expressions for the loss functions used in this work are given by
\begin{gather}\label{eqn:total_loss}
    \mathcal{L} = \textcolor{red}{\sum_{n=1}^{3}\mathcal{L}_{E_n} + \sum_{n>n'}^{3}\mathcal{L}_{\Delta E_{nn'}} + \sum_{n=1}^{3}\mathcal{L}_{\partial E_n/\partial\bm{R}}} + \textcolor{blue}{\sum_{n=1}^{3} \mathcal{L}_{H_{nn}^\text{d}}} + \textcolor{magenta}{\sum_{n>n'}^{3} \mathcal{L}_{\bm{d}_{nn'}}}, 
\end{gather}
where the red, blue, and magenta terms correspond to $\mathcal{L}_\text{core}$, $\mathcal{L}_\text{diab}$, and $\mathcal{L}_\text{nacv}$, respectively, as detailed in the main text. Unless otherwise noted, we use the mean squared error (MSE) defined as $\text{MSE}(\bm{X}) = \sum_j^{M} \frac{1}{M} (X_j^\text{DANN} -X_j^\text{RI-CC2})$ for each quantity. The individual loss components are then
\begin{gather}
    \mathcal{L}_{E_n} = \rho_\text{e} \text{MSE}(E_n), \\
    \mathcal{L}_{\Delta E_{nn'}} = \rho_\text{gap} \text{MSE}(\Delta E_{nn'}), \\
    \mathcal{L}_{\partial E_n/\partial\bm{R}} = \rho_\text{f} \text{MSE}(\partial E_n/\partial\bm{R}), \\
    \mathcal{L}_{H_{nn'}^\text{d}} = \rho_\text{diab} \text{MSE}(H_{nn'}^\text{d}),
\end{gather}
where $\rho_\text{e}$, $\rho_\text{gap}$, $\rho_\text{f}$, and $\rho_\text{diab}$ are scaling factors for the energies, energy gaps, forces, and diabatic Hamiltonian elements, respectively. Special care is required for the nonadiabatic coupling vectors (NACVs) because of the arbitrary sign (phase) of the electronic wavefunctions. Meanwhile, the NACV near conical intersections and avoided crossings might diverge. To handle this ambiguity and divergent behaviors, we adopt the phaseless loss scheme from Ref.~\cite{westermayr_chemsci2019_loss}:
\begin{equation}
    \mathcal{L}_{\bm{d}_{nn'}} = \rho_\text{nacv} \min\left[ \text{MSE}(\bm{b}_{nn'}^\text{DANN}, \bm{b}_{nn'}^\text{RI-CC2}),\text{MSE}(\bm{b}_{nn'}^\text{DANN}, -\bm{b}_{nn'}^\text{RI-CC2})\right],
\end{equation}
where $\bm{b}_{nn'}$ is the energy scaled nonadiabatic coupling vector:
\begin{equation}
    \bm{b}_{nn'} = \bm{d}_{nn'} \Delta E_{nn'} 
\end{equation}
and $\rho_\text{nacv}$ is the scaling factor for the NACV loss.

\begin{figure}[htbp]
    \centering
    \includegraphics[width=0.9\linewidth]{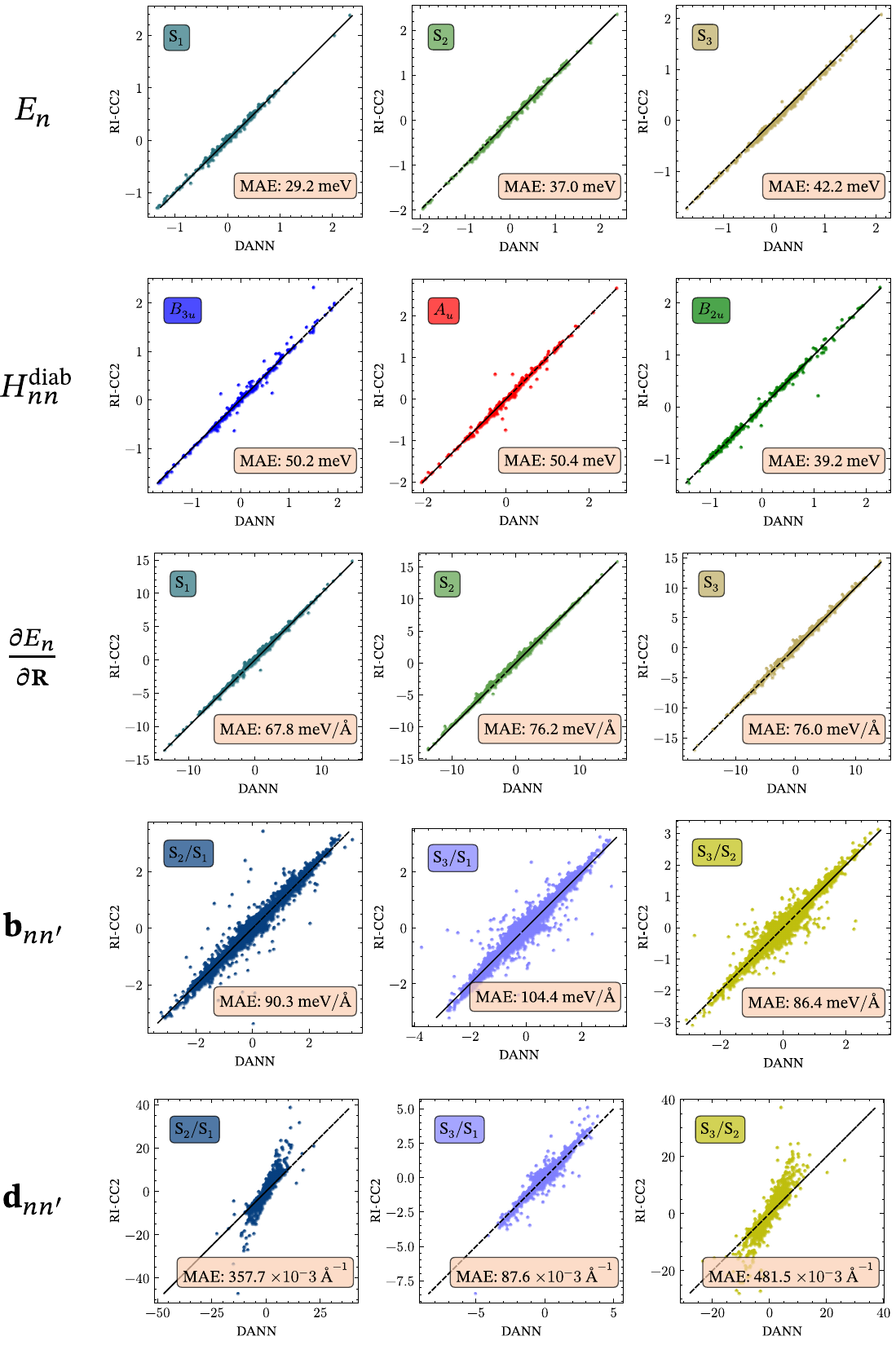}
    \caption{Performance of the DANN model on the validation set with 257 geometries}
    \label{sfig:performance_of_DANN}
\end{figure}

The DANN model implemented in the NeuralForceField package \cite{NeuralForceField} was used. In our production model, 128 features were used to encode both scalar and vectorial (equivariant) quantities, the Swish activation function \cite{ramachandran_arxiv2017_swish} was employed, and 20 radial basis functions (RBFs) were used to construct the rotationally invariant filters. The cutoff radius for constructing the neighbour lists was $5$ Å. Three equivariant message-passing layers are stacked to
propagate local geometric information. The production model was trained with the AdamW optimizer~\cite{loshchilov_arxiv2019_adamw} combined with the Jacobian descent method~\cite{quinton_arxiv_jacobian} to handle the multi-objective loss function. The scaling factors for the individual loss terms are given in Table~\ref{stab:loss_parameters}. A warm restart learning rate schedule~\cite{loshchilov_arxiv2017_SGDR} was used, with the learning rate periodically oscillating between $10^{-3}$ and
$10^{-6}$. The full dataset of 2545 geometries was randomly split into a training set (2288 geometries, ca. 90\%) and a validation set (257 geometries, ca. 10\%). Training was performed for 800 epochs with a batch size of 10, and the model parameters achieving the lowest validation loss were selected as the final model. The performance of this model on the validation set is demonstrated in Figure~\ref{sfig:performance_of_DANN}.

\begin{table}[htbp]
    \centering
    \caption{The weight for each loss components used in this work}
    \label{stab:loss_parameters}
    \begin{tabular}{rrrrr}
    \toprule
    $\rho_\text{e}$ & $\rho_\text{gap}$ & $\rho_\text{f}$ & $\rho_\text{diab}$ & $\rho_\text{nacv}$ \\
    \midrule 
    0.2 & 0.5 & 2.0 & 5.0 & 2.0  \\
    \bottomrule
    \end{tabular}
\end{table}

\clearpage

\clearpage

\section{\emph{Ab initio} predictions of the ultrafast internal conversion time}
\begin{figure}[htbp]
    \centering
    \includegraphics[width=0.80\linewidth]{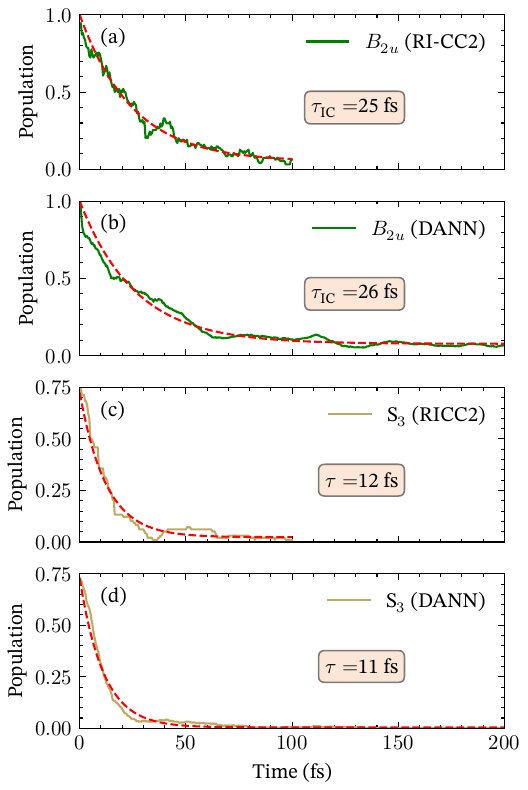}
    \caption{Fitting of the relaxation timescale of the diabatic $B_\text{2u}$ and the adiabatic $\text{S}_3$ states.}
    \label{sfig:taufitting}
\end{figure}


\begin{table}[htbp]
\centering
\caption{\label{stab:abinitio_sufracehopping}Curated predictions of timescales of the ultrafast internal conversion from \emph{ab initio} surface hopping}
\begin{threeparttable}
\begin{tabular}{llll}
\toprule
\multicolumn{2}{l}{Method}                                                        & $\tau_\text{IC} \,(\text{fs})$ & Notes                                                         \\
\midrule
\multirow{2}{*}{Expt.}            & Ref.~\cite{stert2000tdpes_spec}
                                  & $20 \pm 10$
                                  & TRPES\tnote{$\alpha$} 
                                  \\
                                  & Ref.~\cite{horio2009trpei,suzuki_jcp2010_trpei}
                                  & $22 \pm 3$                     
                                  & TRPEI \tnote{$\beta$}
                                  \\
\multirow{2}{*}{TDDFT/B3LYP/TZVP} & Ref.~\cite{werner_cp2008_tddft_b3lyp}
                                  & $21$
                                  & 60 trajectories \tnote{$\dagger$}
                                  \\
                                  & Ref.~\cite{tomasello_jpca2014_surfacehopping}
                                  & $41$
                                  & 100 trajectories \tnote{$\ddagger$}
                                  \\
TDDFT/B3LYP/def2-TZVP             & Ref.~\cite{vogt_jctc2025_semi_extended}
                                  & $10$
                                  & 200 trajectories \tnote{$\dagger$}
                                  \\
TDDFT/B3LYP/cc-pVDZ               & Ref.~\cite{buzsaki_chemrxiv2026}
                                  & $74$
                                  & 76 trajectories \tnote{$\ddagger$}  
                                  \\
TD-DFT/RPBE/cc-pVDZ               & Ref.~\cite{buzsaki_chemrxiv2026}
                                  & $32$
                                  & 50 trajectories \tnote{$\ddagger$}
                                  \\
ADC(2)/aug-cc-pVDZ                & Ref.~\cite{xie2019_assessing}
                                  & $23$
                                  & 500 trajectories \tnote{$\ddagger$}
                                  \\
ADC(2)/aug-cc-pVDZ                & Ref.~\cite{Chang2025}
                                  & $24\pm1.2$
                                  & 170 trajectories \tnote{$\ddagger$}
                                  \\
MS-CASPT2(10e,8o)/cc-pVDZ         & Ref.~\cite{buzsaki_chemrxiv2026}
                                  & $9$
                                  & 55 trajectories \tnote{$\ddagger$}
                                  \\
\multirow{2}{*}{RI-CC2/cc-pVDZ}   & \multirow{2}{*}{\textbf{This work}}
                                  & ${25}$                             
                                  & 98 trajectories (\emph{ab initio}) \tnote{$\ddagger$} \\
                                  &
                                  & ${26}$
                                  & 500 trajectories (DANN) \tnote{$\ddagger$} \\
\bottomrule
\end{tabular}
\begin{tablenotes}
    \item[$\alpha$] Time-resolved photoelectron spectroscopy.
    \item[$\beta$] Time-resolved photoelectron imaging.
    \item[$\dagger$] $\tau_\text{IC}$ fitted from adiabatic population.
    \item[$\ddagger$] $\tau_\text{IC}$ fitted from diabatic population.
\end{tablenotes}
\end{threeparttable}
\end{table}

\clearpage

\section{Additional time evolution of the nuclear density data}
\begin{figure}[htbp]
    \centering
    \includegraphics[width=0.9\linewidth]{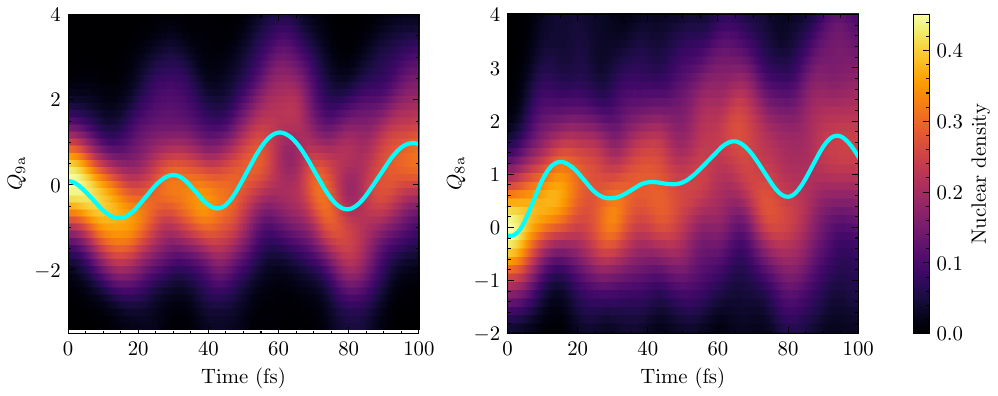}
    \caption{Time evolution of the nuclear density projected onto the $Q_\text{9a}$ (a) and $Q_\text{8a}$ (b) modes from the RI-CC2 simulations. The color map represents the nuclear density, while the cyan line shows the average value of the $Q_\text{9a}$ and $Q_\text{8a}$ coordinates as a function of time.}
    \label{sfig:heatmap_nuc_ricc2}
\end{figure}

\clearpage
\begin{figure}[htbp]
    \centering
    \includegraphics[width=0.9\linewidth]{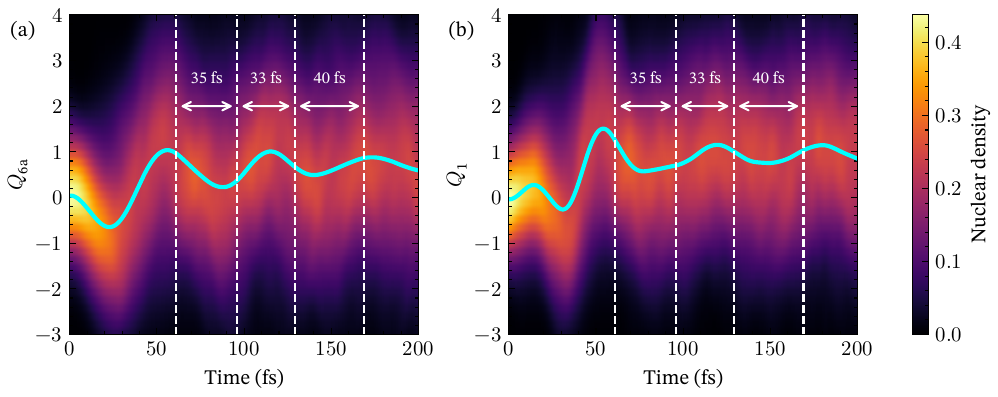}
    \caption{Time evolution of the nuclear density projected onto the $Q_\text{6a}$ (a) and $Q_\text{1}$ (b) modes from the DANN simulations. The color map represents the nuclear density, while the cyan line shows the average value of the $Q_\text{6a}$ and $Q_\text{1}$ coordinates as a function of time.}
    \label{sfig:heatmap_nuc_mismatch}
\end{figure}

\clearpage
\section{{Confidence Intervals for Ensemble Averaging}\label{sec:confidence}}

\begin{figure}[htbp]
    \centering
    \includegraphics[width=0.75\linewidth]{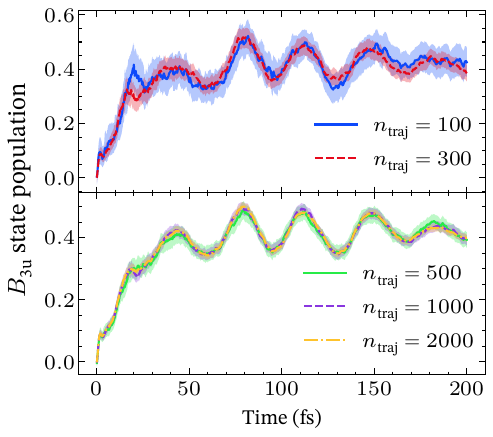}
    \caption{Population of the B$_\text{3u}$ state as a function of time for different ensemble sizes. The lines represent the averaged populations, and the shaded areas denote the 99\% confidence intervals estimated via bootstrap resampling \label{sfig:bootstrap}}
\end{figure}

Figure~\ref{sfig:bootstrap} shows the time-dependent population of the B$_\text{3u}$ state obtained from trajectory ensembles of varying sizes, ranging from 100 to 2000 trajectories. The results demonstrate a substantial narrowing of the 99\% confidence intervals when increasing the ensemble size from 100 to 500 trajectories. Further increasing the ensemble size from 500 to 2000 trajectories yields only marginal improvements in the confidence interval width. These observations indicate that 500 trajectories are sufficient to produce smooth and statistically converged population dynamics. We therefore selected 500 trajectories for all on-the-fly simulations presented in the main text.

\begin{figure}[htbp]
    \centering
    \includegraphics[width=0.95\linewidth]{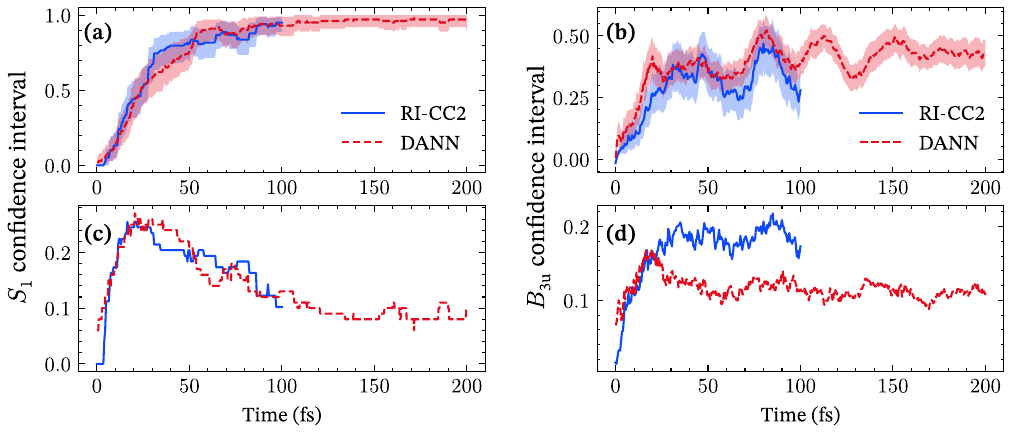}
    \caption{Comparison of the \emph{ab initio} RI-CC2 ensemble (98 trajectories) and a matched 100-trajectory DANN ensemble. (a), (b) Ensemble-averaged S$_1$ and diabatic B$_{3u}$ populations; shaded areas denote the 99\% confidence intervals from bootstrap resampling (2{,}000 resamples with replacement). (c), (d) Corresponding confidence-interval widths as functions of time.\label{sfig:bootstrap_2}}
\end{figure}

To test whether the DANN dynamics faithfully represent the underlying CC2 dynamics,  we randomly selected a 100-trajectory DANN ensemble matched to the 98-trajectory \emph{ab initio} RI-CC2 ensemble and applied the same bootstrap procedure as in Figure~\ref{sfig:bootstrap} (2000 resamples with replacement, 99\% confidence). Figure~\ref{sfig:bootstrap_2}(a) shows excellent agreement for the adiabatic population: the ensemble averages essentially coincide, and their confidence intervals have comparable widths throughout the propagation. For the diabatic population, however, the DANN and RI-CC2 results differ more visibly [Figure~\ref{sfig:bootstrap_2}(b),(d)], with RI-CC2 showing a broader confidence interval. 

Since the adiabatic statistics of the two ensembles coincide, the additional roughness of the RI-CC2 diabatic populations must originate in the diabatization procedure rather than in trajectory statistics. Nevertheless, the majority of shaded RI-CC2 regions essentially overlap those of DANN, meaning that the mean diabatic populations are statistically indistinguishable within the 99\% confidence intervals. The smoother averaged DANN populations are expected by construction: the neural network learns a smoothed representation of the potential-energy surfaces and couplings and cannot reproduce sharp local features of the \emph{ab initio} data.

\begin{figure}[htbp]
    \centering
    \includegraphics[width=0.95\linewidth]{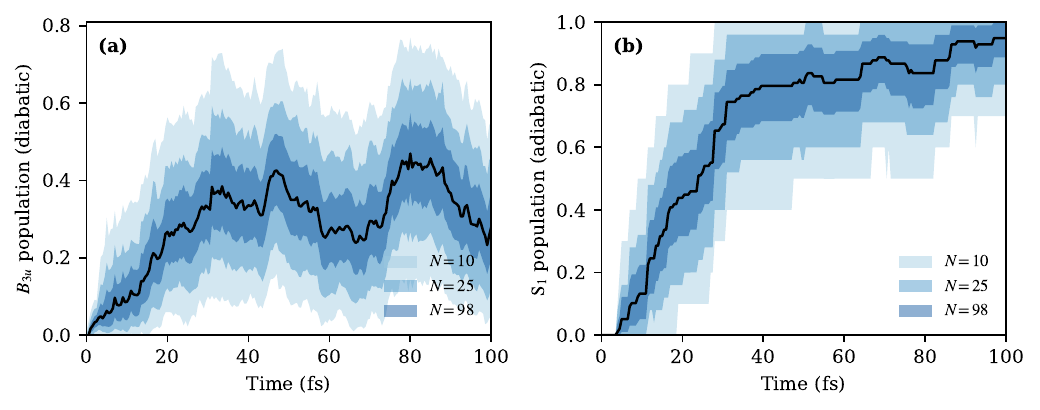}
    \caption{Convergence of the ensemble-averaged B$_{3u}$ population of the \emph{ab initio} RI-CC2 dynamics with the number of trajectories $N$. Black line: the ensemble mean. Shaded areas are the 99\% confidence intervals obtained by resampling $N = 10$, 25, and 98 trajectories. Panels (a) and (b) show the diabatic and adiabatic populations, respectively.\label{sfig:bootstrap_3}}
\end{figure}

Figure~\ref{sfig:bootstrap_3} shows the convergence of the RI-CC2 confidence intervals with the number of trajectories $N$. The width of the 99\% confidence interval follows the expected $N^{-1/2}$ behavior: for the diabatic B$_{3u}$ population, the averaged width decreases from 0.51 ($N = 10$) to 0.33 ($N = 25$) and then to 0.17 ($N = 98$); the adiabatic widths behave analogously (0.50, 0.35, 0.18). Thus, the finite-trajectory uncertainty is comparable for the diabatic and adiabatic populations.

\clearpage
\section{{Comparison between the VC Model and the On-the-Fly Simulations}}

\begin{figure}[htbp]
    \centering
    \includegraphics[width=0.99\linewidth]{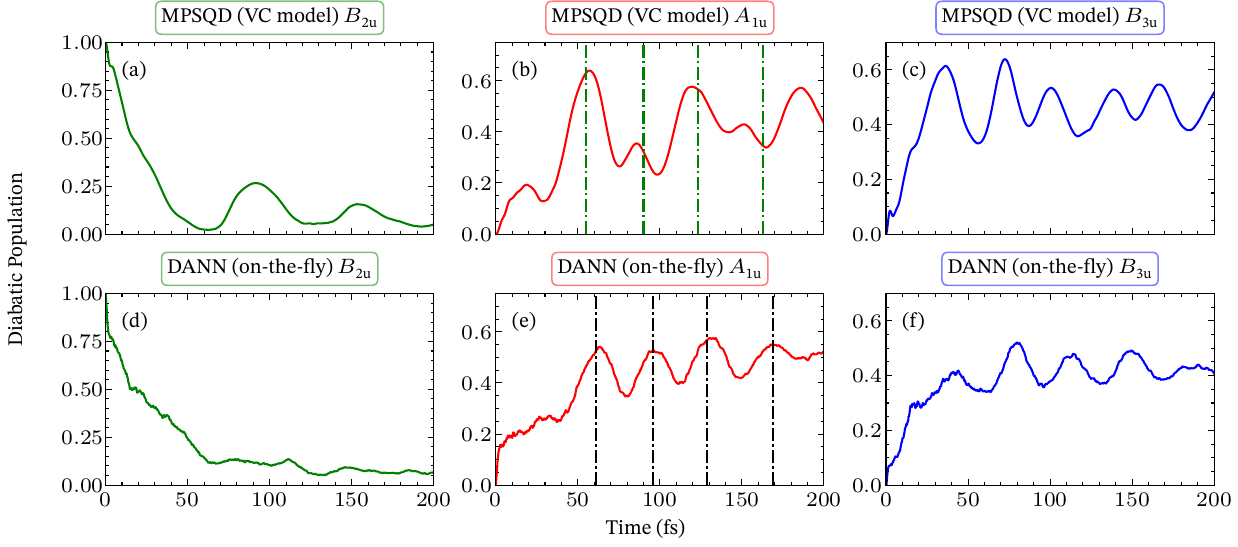}
    \caption{Comparison of the population dynamics obtained from the VC model solved with MPSQD and the on-the-fly surface-hopping dynamics with DANN. Panels (a), (b), and (c) show the populations of the B$_\text{2u}$, A$_\text{1u}$, and B$_{3u}$ diabatic states, respectively, from the VC model. Panels (d), (e), and (f) show the corresponding populations from the on-the-fly dynamics.}
    \label{sfig:direct_comparison_pop}
\end{figure}

Figure~\ref{sfig:direct_comparison_pop} demonstrates that the population dynamics obtained from the VC model and the on-the-fly simulations are in good qualitative agreement. However, two notable differences are observed. First, the VC model predicts a pronounced recurrence of population to the initial B$_{2u}$ state around 90 and 150 fs, which is strongly suppressed in the full-dimensional on-the-fly dynamics. Second, both methods exhibit sustained A$_{1u}$/B$_{3u}$ population oscillations, but with different structures. In particular, the VC model shows a two-component beating (a dominant $\approx$60 fs component and a minor $\approx$30 fs component) for the A$_{1u}$ population, whereas no such structure is apparent in the on-the-fly data. The oscillatory pattern in the VC model occurs slightly earlier than that in the on-the-fly simulations.

These minor discrepancies are expected, since the on-the-fly dynamics explicitly couples the three lowest adiabatic excited states, whereas the VC model is restricted to a reduced subspace of relevant diabatic states.

\begin{figure}[htbp]
    \centering
    \includegraphics[width=0.99\linewidth]{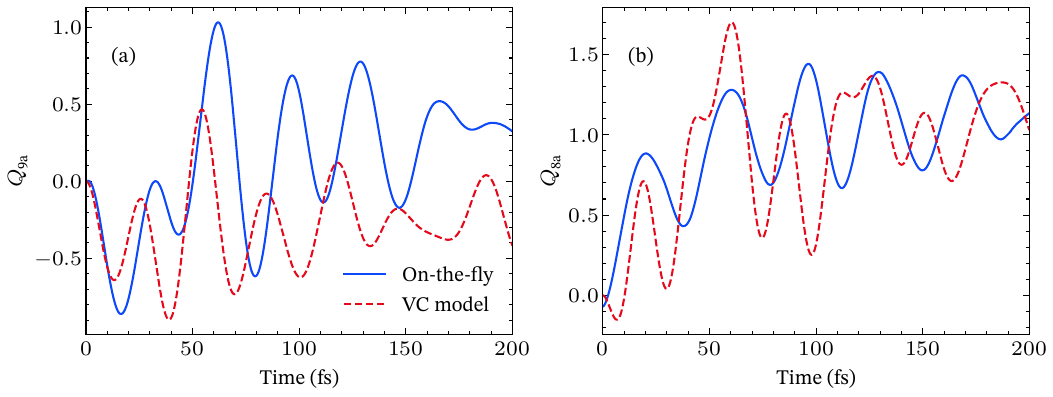}
    \caption{Comparison of the time evolution of the averaged nuclear coordinate obtained from the VC model solved with MPSQD and the on-the-fly surface-hopping dynamics with DANN. Panels (a) and (b) show the dynamics of the $Q_\text{9a}$ and $Q_\text{8a}$ modes, respectively.}
    \label{sfig:direct_comparison_nuc}
\end{figure}

Figure~\ref{sfig:direct_comparison_nuc} presents a direct comparison of the average nuclear dynamics for the Q$_{9a}$ and Q$_{8a}$ modes between the VC model and the on-the-fly simulations. Mirroring the population dynamics, the VC model exhibits an earlier onset of correlated vibrational motion associated with the A$_{1u}$--B$_{3u}$ population transfer. For the Q$_{8a}$ mode, both approaches converge to a similar average displacement of approximately
+1 at 200 fs, which is between the equilibrium positions of the A$_{1u}$ and B$_{3u}$ states as shown in Figure~1d of the main text. In contrast, the averaged Q$_{9a}$ displacement differs markedly: the VC model yields values between $-0.5$ and $0$, whereas the on-the-fly dynamics gives values between $0$ and $0.7$ at 200 fs. This disparity indicates that the full-dimensional dynamics imposes a net positive displacement along Q$_{9a}$, effectively steering the nuclear wavepacket closer toward the A$_{1u}$/B$_{3u}$ crossing seam.

\begin{figure}[htbp]
    \centering 
    \includegraphics[width=0.95\linewidth]{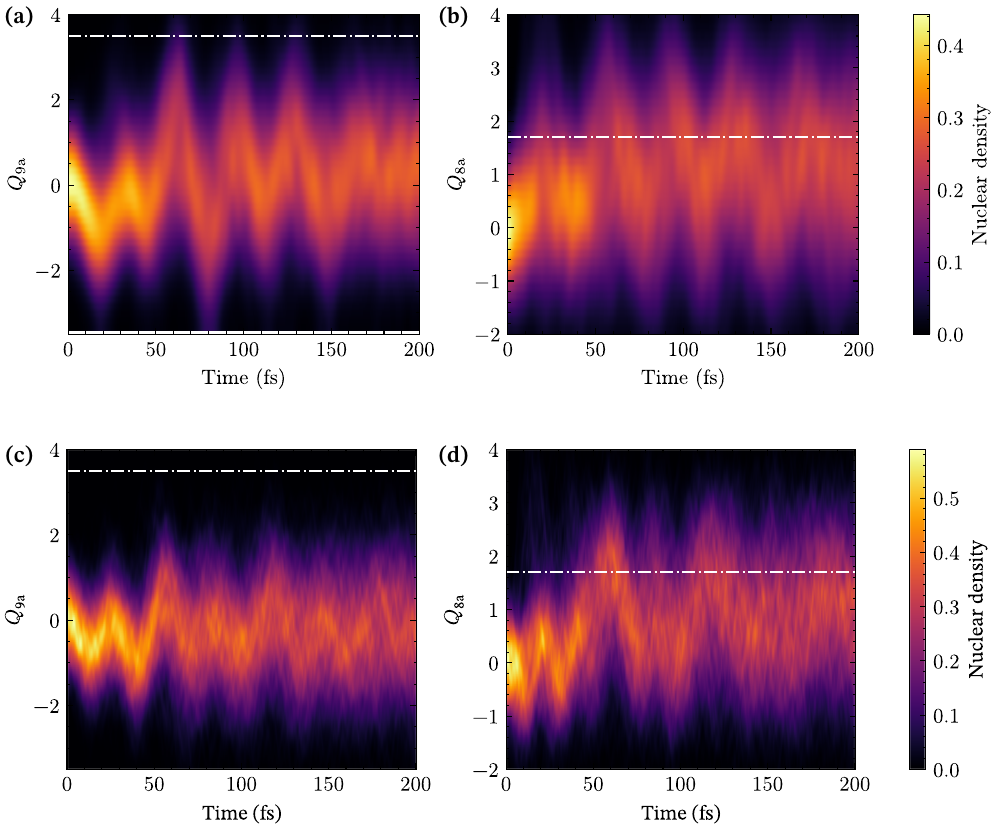}
    \caption{Comparison of the time evolution of the nuclear density obtained from the VC model solved with MPSQD and the on-the-fly surface-hopping dynamics with DANN. Panels (a) and (b) show the dynamics of the $Q_\text{9a}$ and $Q_\text{8a}$ modes, respectively, for the on-the-fly dynamics. Panels (c) and (d) plot the results for the VC model. The dash-dot white line marks the location of the $\text{A}_{1u}/\text{B}_\text{3u}$ seam along the $Q_\text{9a}$ and $Q_\text{8a}$.}
    \label{sfig:vc_otf_nuclear_density}
\end{figure}

Beyond the averaged positions, Figure~\ref{sfig:vc_otf_nuclear_density} shows the time evolution of nuclear densities. In particular, the comparison of panels (a) and (c) shows that the on-the-fly trajectories spread over a much wider region along the $Q_\text{9a}$ coordinate ($-2.5<Q_\text{9a}<3.5$), whereas the VC model predicts a more contracted $Q_\text{9a}$ distribution ($-2.5<Q_\text{9a}<2.5$). In contrast, panels (b) and (d) show that the spread of the $Q_\text{8a}$ mode during the dynamics is comparable between the on-the-fly dynamics and the VC model, which only slightly underestimates the $Q_\text{8a}$ spread. As shown in the following Sec.~\ref{ssec:hop}, the $Q_\text{8a}$ and $Q_\text{9a}$ modes form a pair of totally symmetric tuning coordinates that modulate the $\text{A}_\text{1u}/\text{B}_\text{3u}$ character of the $S_2$ and $S_1$ adiabatic states. Therefore, although the VC model describes the $Q_\text{8a}$ tuning motion comparably to the on-the-fly dynamics, it does not fully capture the tuning contribution of $Q_\text{9a}$ to the nonadiabatic dynamics of pyrazine.

\clearpage
\section{{State ordering: RI-CC2 vs XMCQDPT2}}
\begin{figure}[htbp]
    \centering
    \includegraphics[width=0.95\linewidth]{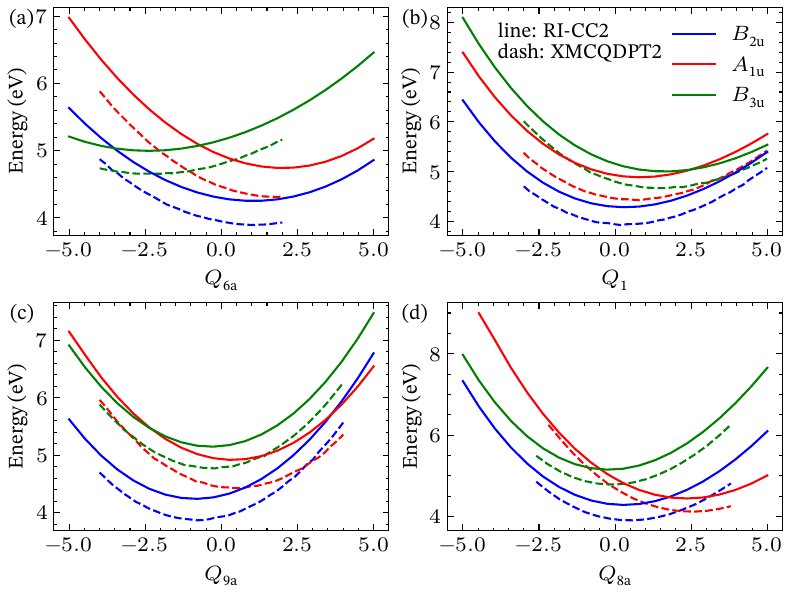}
    \caption{Potential energy profiles along the four most significant tuning modes: (a) $Q_{\text{6a}}$, (b) $Q_{\text{1}}$, (c) $Q_{\text{9a}}$, and (d) $Q_{\text{8a}}$. The green, red, and blue curves correspond to the diabatic $B_{\text{3u}}$, $A_{\text{1u}}$, and $B_{\text{2u}}$ states, respectively. Solid and dashed lines denote the RI-CC2 and XMCQDPT2 results, respectively. Adapted with permission from Ref.~\cite{sala_pccp_qd2015}. Copyright 2015 the PCCP Owner Societies.}
    \label{sfig:state_ordering_compare}
\end{figure}
Figure~\ref{sfig:state_ordering_compare} shows one-dimensional potential energy profiles along the four key tuning modes. While RI-CC2 consistently yields higher absolute excitation energies than XMCQDPT2, the relative state ordering and the positions of the crossing seams are largely preserved between the two methods. Furthermore, the RI-CC2 population dynamics agree well with the ADC(2) populations compared directly in Sec.~S12.

\section{{Population dynamics: RI-CC2 vs ADC(2)}}
\begin{figure}[htbp]
    \centering
    \includegraphics[width=0.95\linewidth]{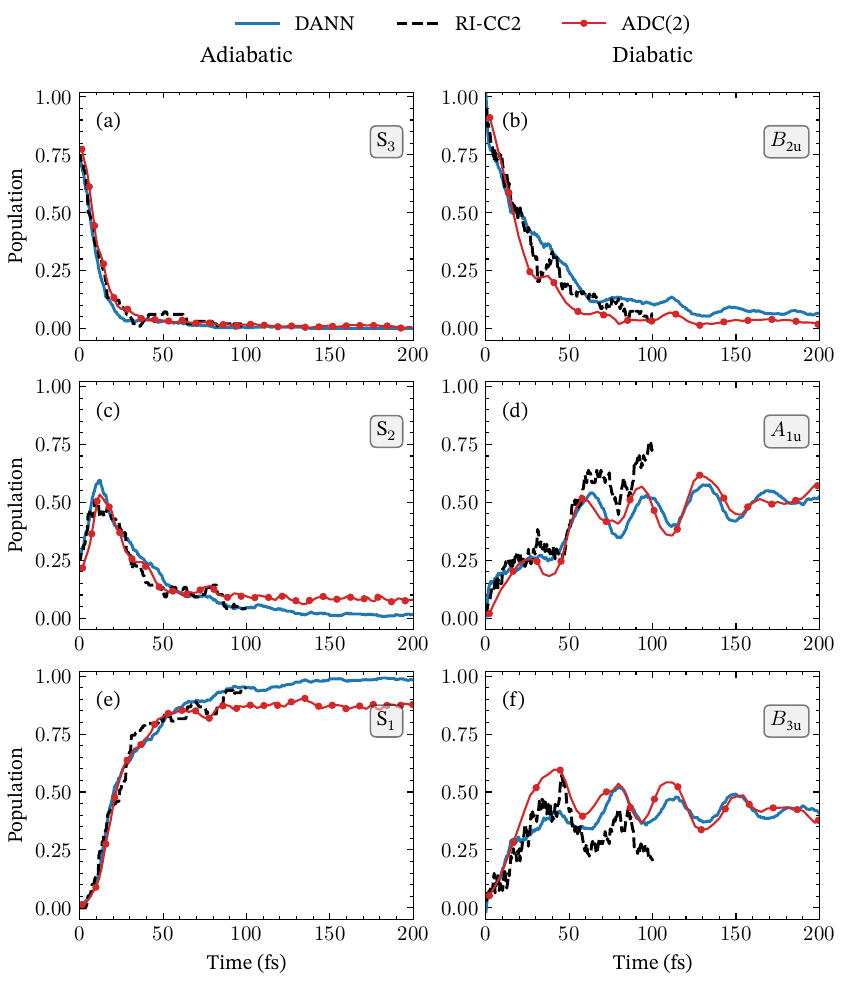}
    \caption{Comparison of the ensemble-averaged adiabatic (a, c, e) and diabatic (b, d, f) populations of pyrazine obtained from on-the-fly surface-hopping dynamics at the RI-CC2 level (black dashed, 0–100 fs), with the DANN model (blue, 0–200 fs), and with the published ADC(2) dynamics of Ref.~\cite{xie2019_assessing} (red, digitized, 0--200 fs). The red ADC(2) data are adapted with permission from Ref.~\cite{xie2019_assessing}. Copyright 2019 AIP Publishing.}
    \label{sfig:adc2_comparison}
\end{figure}

Figure~\ref{sfig:adc2_comparison} compares the ensemble-averaged adiabatic [panels (a), (c), (e)] and diabatic [panels (b), (d), (f)] populations from the RI-CC2 and the DANN simulations with the published ADC(2) dynamics of Xie \emph{et al.}~\cite{xie2019_assessing}, digitized from the original publication.

All three simulations follow the same relaxation pathway. The adiabatic $\text{S}_3$ population decays within 30~fs, with $\text{S}_1$ growing on the same timescale. The S$_2$ population, partially populated at initialization, first increases within 10~fs and is then rapidly depleted. RI-CC2 and ADC(2) show excellent agreement for the $\text{S}_3$ decay, whereas the long-time plateaus differ for $\text{S}_2$ and $\text{S}_1$. The CC2-based simulations predict an almost fully depleted $\text{S}_2$ population, while ADC(2) retains a residual $\text{S}_2$ population of 0.1 for $t>120$~fs. As for the diabatic populations, all methods predict a monotonic $B_\text{2u}$ decay, with ADC(2) predicting a slightly faster decay (Table~\ref{stab:abinitio_sufracehopping}). After the initial population increase of the $A_\text{1u}$ and $B_\text{3u}$ states ($t<40$~fs), both exhibit sustained oscillations up to 200~fs.

This agreement is notable because the two methods are formally closely related yet distinct. In particular, CC2 and ADC(2) excitation energies agree to second order, but the CC2 Jacobian is non-Hermitian. Near same-symmetry conical intersections, CC2 can therefore produce complex-conjugate excitation energies (see main text). In contrast, ADC(2) is Hermitian, so its roots remain real and it preserves the correct conical-intersection topology. Despite these differences,
Figure~\ref{sfig:adc2_comparison} shows that the two methods reach semi-quantitatively agreement in the ensemble-averaged population dynamics for pyrazine.

\clearpage
\section{{Hop-event analysis}\label{ssec:hop}}

    To support our interpretation of the population dynamics, we analyzed all surface-hopping events in the trajectory ensemble. For each hop, we extracted the adiabatic states involved, the energy gap at the hop geometry, and the projection of the hop geometry onto the Q$_\text{9a}$ and Q$_\text{8a}$ modes. Note that in the main text, we used a 500-trajectory ensemble. The analysis in this section uses the full 2000-trajectory ensemble.

\begin{figure}[htbp]
    \centering
    \includegraphics[width=0.95\linewidth]{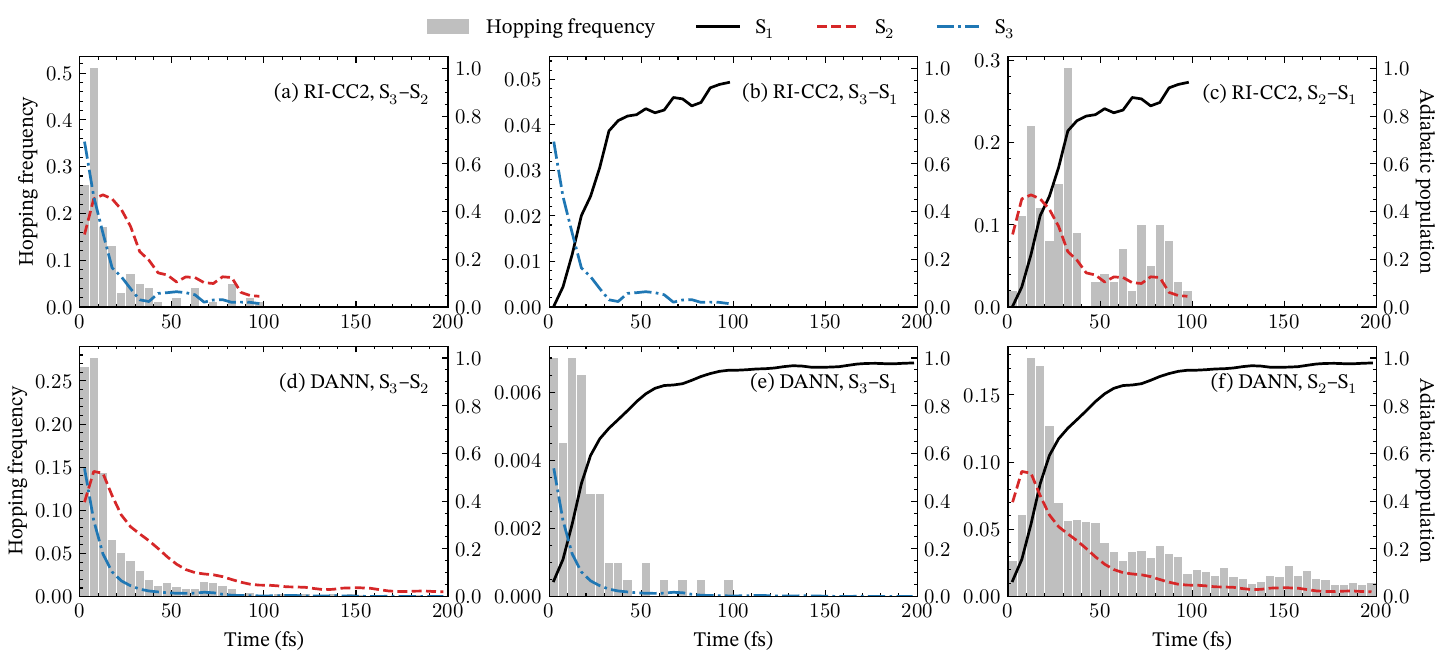}
    \caption{Hopping frequency (bars; number of hops between the indicated pair of states per 5 fs interval, normalized by the total number of trajectories) and the fraction of trajectories on the involved adiabatic states (lines) as functions of time. Top row: RI-CC2 ensemble; bottom row: DANN ensemble. Panels (a)/(d): S$_3$--S$_2$ hops; (b)/(e): S$_3$--S$_1$ hops; (c)/(f): S$_2$--S$_1$ hops.}
    \label{sfig:hop_rate}
\end{figure}

Figure~\ref{sfig:hop_rate} shows the time evolution of the adiabatic populations and the hopping frequency for each pair of states. Hopping is dominated by the sequential pathway S$_3$ $\to$ S$_2$ $\to$ S$_1$. Direct S$_3$--S$_1$ hops are absent from the smaller RI-CC2 ensemble and account for only 1.4\% of the DANN hops. Most hopping events occur within the first 60~fs, the window in which the initially populated $\text{S}_3$ and $\text{S}_2$ states decay. During this window, the S$_1$ population rises to approximately 0.9 for both RI-CC2 and DANN. In total, only 60 of the 303 RI-CC2 hops and 1243 of the 5056 DANN hops occur after 60~fs, corresponding to about 0.6 hops per trajectory in the late-time window. This indicates that the sustained A$_{1u}$/$B_{3u}$ population oscillations shown in Figure 5 of the main text reflect oscillations of the diabatic character on the dominant S$_1$ adiabatic surface rather than repeated S$_1$/$S_2$ hops.

\begin{figure}[htbp]
    \centering
    \includegraphics[width=0.95\linewidth]{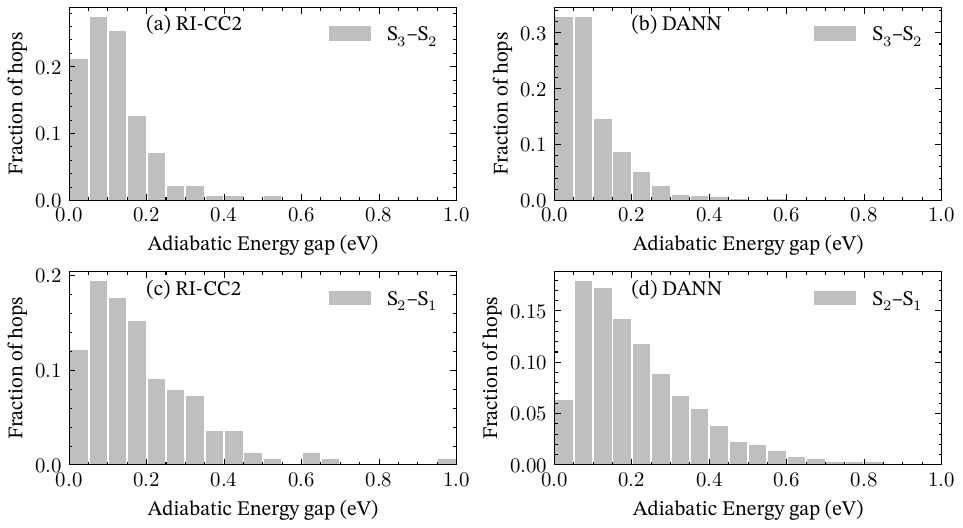}
    \caption{Distribution of the adiabatic energy gap at the hop events of the indicated state pairs, normalized to the total number of hops of each pair: S$_3$--S$_2$ hops of RI-CC2 (a) and DANN (b); S$_2$--S$_1$ hops of RI-CC2 (c) and DANN (d). The hops occur preferentially at small gaps.}
    \label{sfig:gap_analysis}
\end{figure}

Figure~\ref{sfig:gap_analysis} shows that hops for both state pairs occur preferentially at small gaps. Few hops occur when the gap exceeds 0.4~eV for the $\text{S}_3-\text{S}_2$ transition and 0.6~eV for the $\text{S}_2-\text{S}_1$ transition, consistent with the surface-hopping picture. The distribution of hop fractions resembles that reported by Xie \emph{et al.}\cite{xie2019_assessing} using ADC(2) (see their FIG. 6).
\begin{figure}[htbp]
    \centering
    \includegraphics[width=0.95\linewidth]{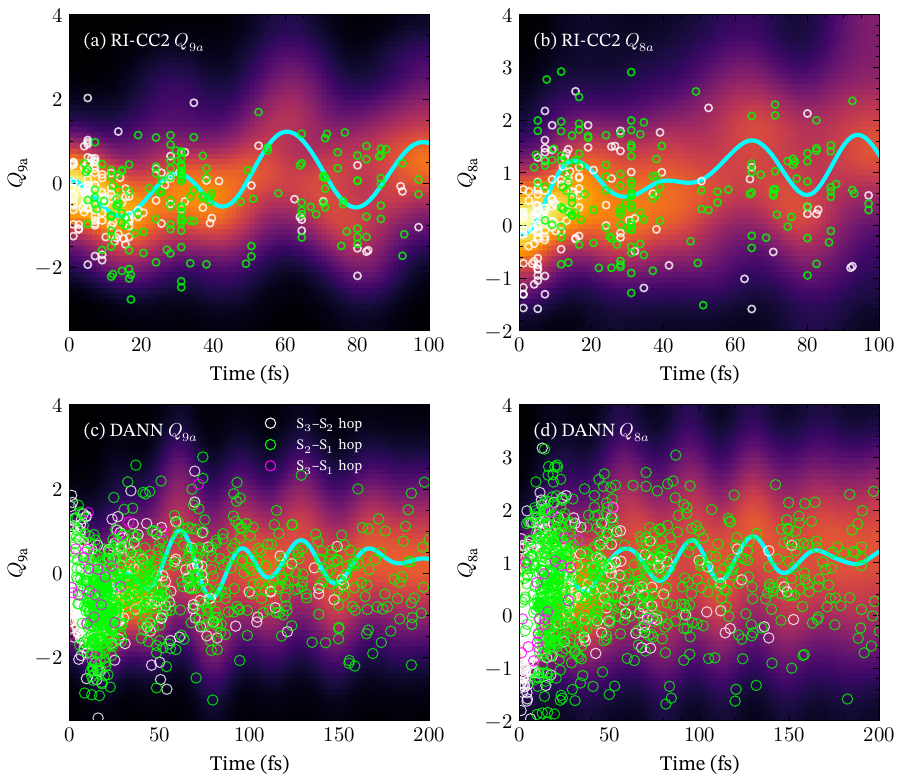}
    \caption{Hop events overlaid on the Q$_\text{9a}$ (a, c) and Q$_\text{8a}$ (b, d) nuclear-density maps for the RI-CC2 (a, b) and DANN (c, d) ensembles. The cyan lines show the ensemble-averaged Q$_\text{9a}$ and Q$_\text{8a}$ coordinates. White, magenta and green circles denote instances of S$_3$--S$_2$, S$_3$--S$_1$, and S$_2$--S$_1$ transitions, respectively.}
    \label{sfig:hops_on_density}
\end{figure}

Figure~\ref{sfig:hops_on_density} overlays the hopping events on the Q$_\text{9a}$ and Q$_\text{8a}$ nuclear-density maps. The white circles indicate $\text{S}_3-\text{S}_2$ transitions, which occur mainly at $t<60$~fs, in agreement with Figure~\ref{sfig:hop_rate}. The green circles indicate the subsequent $\text{S}_2-\text{S}_1$ transitions. The hopping events do not cluster at the $Q_\text{9a} \approx 3.5$ or $Q_\text{8a} \approx 1.8$ crossing regions. This indicates that both Q$_\text{9a}$ and Q$_\text{8a}$ act as tuning coordinates that modulate the A$_{1u}$/$B_{3u}$ character along the dynamics rather than as drivers of the population transfer.

\clearpage
\section{Stability analysis of the RI-CC2 eigensolver}
\label{sec:stability_audit}

\begin{figure}[htbp]
    \centering
    \includegraphics[width=0.95\linewidth]{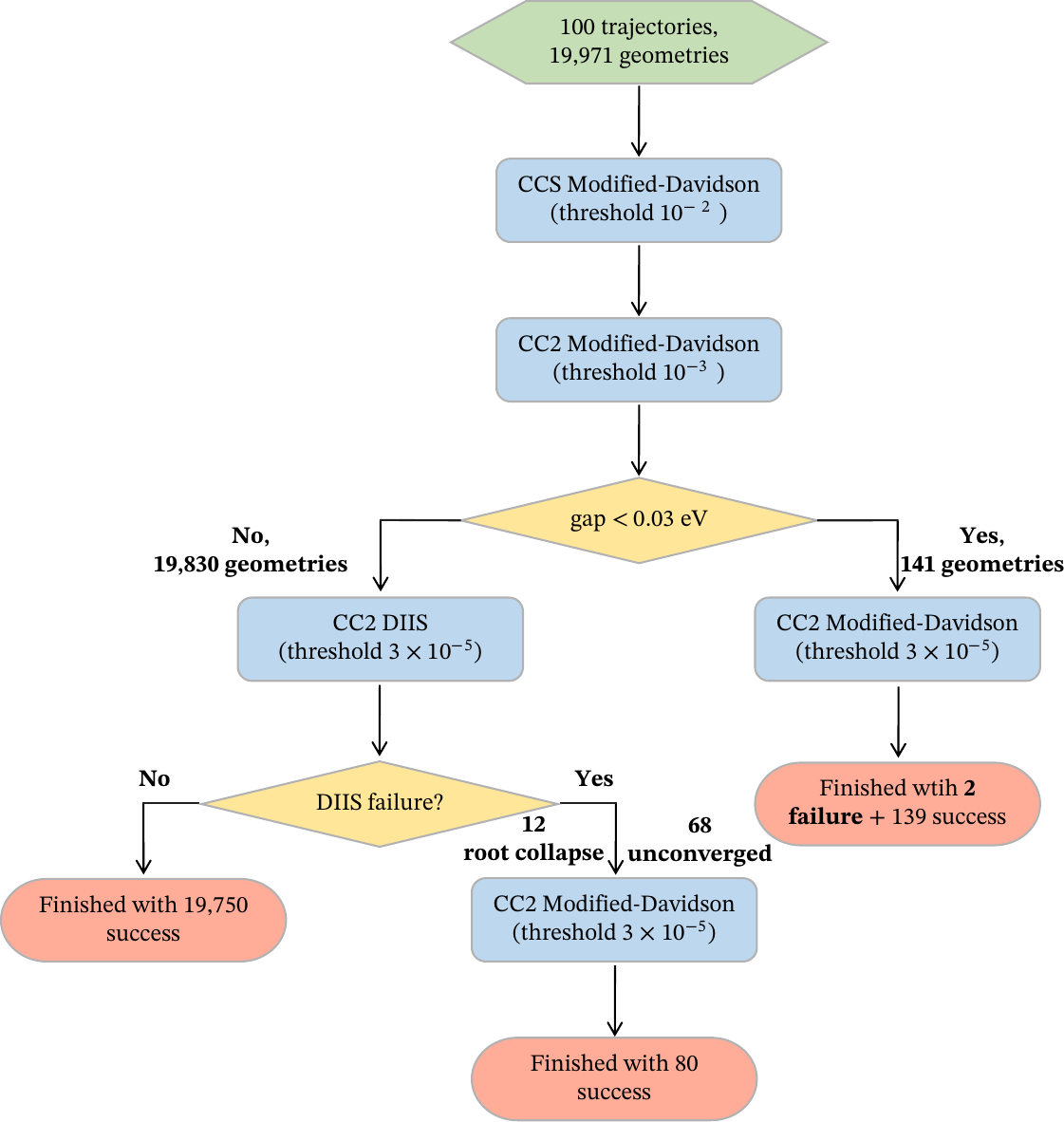}
    \caption{Flowchart of the RI-CC2 eigensolver procedure and the outcome of the stability analysis for all 19,971 quantum-chemistry calculations in the 100-trajectory ensemble.}
    \label{sfig:stability_analysis}
\end{figure}

To assess the numerical stability of the RI-CC2 eigensolver during the nonadiabatic dynamics, we analyzed the convergence behavior of every quantum-chemistry calculation along all 100 \emph{ab initio} trajectories. In total, 19,971 molecular geometries were evaluated. The complete decision workflow, including the convergence thresholds and fallback strategy, is summarized in Figure~\ref{sfig:stability_analysis}.

At each geometry, the excitation energies and amplitudes were first pre-optimized using the modified-Davidson solver\cite{hattig_2000_moddav}, with convergence thresholds of $10^{-2}$ for CCS and $10^{-3}$ for CC2. The resulting solutions were subsequently refined to a tight residual threshold of $3\times10^{-5}$. If the pre-optimization identified a near-degenerate pair of excitation energies with an energy gap smaller than $0.03$~eV, the calculation was directly converged using the modified-Davidson solver rather than DIIS. This criterion was triggered for 141 geometries. Among these cases, 139 geometries completed the step, while two geometries failed to converge within 200 iterations, terminating the corresponding trajectories at $t=63.0$ and $72.5$~fs. Among the 139 completed geometries, two reached exactly degenerate real eigenvalues (gap $< 10^{-6}$~eV). Independent \emph{ricc2} (Turbomole) calculations at the two failing geometries confirmed that the coalescing roots form genuine complex-conjugate pairs of the CC2 Jacobian.

For the remaining 19,830 geometries, the excitation-energy gap exceeded $0.03$~eV after pre-optimization, and the tight convergence step was initially performed using DIIS. The DIIS procedure successfully converged for 19,750 geometries. For the remaining 80 geometries, DIIS either collapsed onto an incorrect electronic root (12 geometries) or failed to converge within the iteration limit (68 geometries). In the former case, the DIIS solution for one excited state drifted onto that of an adjacent state. These cases were automatically redirected to the modified-Davidson solver, which successfully converged all 80 calculations.

Overall, the stability analysis demonstrates that the combined DIIS/modified-Davidson strategy converged 19,969 out of 19,971 quantum-chemistry calculations. The two unsuccessful cases occurred on trajectories approaching regions of near-degeneracy and led to trajectory termination; both trajectories were excluded from all statistical analyses. Finally, we note that the modified-Davidson solver only \emph{approximately} solves the nonlinear, partitioned CC2 Jacobian. Nevertheless, in practice, we find that the excitation energies obtained with the modified-Davidson solver generally differ from those obtained with DIIS by much less than 0.01~eV. We therefore consider the modified-Davidson solver sufficiently accurate to serve as a fallback when DIIS fails or when the excitation-energy gap becomes small ($<0.03$~eV).

\section{Dataset and DANN model for pyrazine}
The complete dataset of this work is publicly available in the Figshare repository: 

\href{https://doi.org/10.6084/m9.figshare.31866094}{DOI: 10.6084/m9.figshare.31866094}.

\clearpage

\bibliography{si}